\documentclass[electronics,article,accept,pdftex,moreauthors]{Definitions/mdpi} 

\usepackage{makecell} 
\usepackage{epstopdf}
\usepackage{verbatim}
\usepackage{mathrsfs}
\usepackage{graphicx}
\newcolumntype{H}{>{\setbox0=\hbox\bgroup}c<{\egroup}@{}}

\firstpage{1} 
\pubvolume{1}
\issuenum{1}
\articlenumber{0}
\pubyear{2026}
\copyrightyear{2026}
\externaleditor{Enzo Pasquale Scilingo and Gianluca Rho}
\datereceived{3 February 2026} 
\daterevised{6 March 2026} 
\dateaccepted{10 March 2026 } 
\datepublished{} 
\Title{Unmasking Biases and Reliability Concerns in Convolutional Neural Networks Analysis of Cancer Pathology Images}

\Author{Michael Okonoda %MDPI: Please carefully check the accuracy of names and affiliations. Please notice that any changes about authorship, affiliations, and corresponding authors are not allowed after the paper has been accepted, including adding, removing, or changing the order.  Author response: The name is correct.
, Eder Martinez, Abhilekha Dalal and Lior Shamir *\orcidA{} %MDPI: We added the * here to indicate corresponding author, please confirm.   Author response: The corresponding author is correct.
}  

\AuthorNames{Michael Okonoda, Eder Martinez, Abhilekha Dalal, Lior Shamir}
\address[1]{%
Department %MDPI: The affiliation numbers should appear in numerical order. Please check the suggested changes.   Author response:  I am not entirely sure I see a change but it looks okay. No problem.
 of Computer Science, Kansas State University, Manhattan, KS 66506, USA%MDPI: We added country name, please confirm.
; \linebreak  mokonoda@ksu.edu (M.O.); edermar@ksu.edu (E.M.); adalal@ksu.edu (A.D.) %MDPI: We added the email addresses here according to those submitted online at susy.mdpi.com. Please confirm.  Author response: Thank you. That looks okay.
}

\corres{\hangafter=1 \hangindent=1.05em \hspace{-0.82em}Correspondence: lshamir@mtu.edu}

\abstract{
Convolutional Neural Networks have shown promising effectiveness in identifying different types of cancer from radiographs. However, the opaque nature of CNNs makes it difficult to fully understand the way they operate, limiting their assessment to empirical evaluation. Here we study the soundness of the standard practices by which CNNs are evaluated for the purpose of cancer pathology. Thirteen highly used cancer benchmark datasets were analyzed, using four common CNN architectures and different types of cancer, such as melanoma, carcinoma, colorectal cancer, and lung cancer. We compared the accuracy of each model with that of datasets made of cropped segments from the background of the original images that do not contain clinically relevant content. Because the rendered datasets contain no clinical information, the null hypothesis is that the CNNs should provide mere chance-based accuracy when classifying these datasets. The results show that the CNN models provided high accuracy when using the cropped segments, sometimes as high as 93\%, even though they lacked biomedical information. These results show that some CNN architectures are more sensitive to bias than others. The analysis shows that the common practices of machine learning evaluation might lead to unreliable results when applied to cancer pathology. These biases are very difficult to identify, and might mislead researchers as they use available benchmark datasets to test the efficacy of CNN~methods. 
}

\keyword{bias; deep learning; pathology; robustness}

\begin{document}

\section{Introduction}
\label{introduction}

\textls[-15]{Pathology benchmark datasets %\citep{barisoni2020digital} 
play a critical role in advancing automatic diagnostic systems through the use of deep learning and other machine learning models~\citep{echle2021deep,zhu2016deep,orlov2010automatic, xiao2024convolutional,xu2025edge,ciga2022self,de2021machine,rajadurai2024precisionlymphonet,chen2024can,shen2017deep}. 
These biomedical images can be used to train machine learning systems to identify cancer through different types of imaging modalities, such as X-rays~\citep{irede2024medical}, Magnetic Resonance Imaging (MRI)~\citep{weigel2015extended,haris2015molecular}, Positron Emission Tomography (PET)~\citep{schwenck2023advances}, Computed Tomography (CT)~\citep{reuveni2011targeted, ahmed2020lung}, Optical Coherence Tomography (OCT)~\citep{schwartz2022ovarian}, ultrasound~\citep{ayana2021transfer}, and endoscopy~\citep{islam2022deep}.}

Artificial Intelligence (AI), and specifically deep neural networks, have been advancing rapidly, with a significant impact on various scientific domains. Convolutional neural networks~\citep{lecun2015deep,gu2018recent} have become the dominant approach to image recognition and analysis tasks. Their ability to learn features directly from the data has led to significant advancements in various fields related to computer vision, including automated image-based pathology ~\citep{yamashita2018convolutional}.

% CNNs have exhibited their efficacy in addressing intricate challenges within the domain of computer vision over the past two decades. They combine accessibility and ease of use with relatively high performance. 

%But despite their numerous strengths, CNNs also possess notable drawbacks. % The opaqueness of its feature selection process gives it its ``Black-Box" nature, which can introduce concerns regarding its reliability and interpretability.  

%In recent years, CNNs have had a far-reaching impact on biomedical image analysis, advancing abilities in medical diagnostics, prognostics, and therapeutic guidance~\citep{shen2017deep,echle2021deep,zhu2016deep,orlov2010automatic,xiao2024convolutional,xu2025edge,ciga2022self,de2021machine,rajadurai2024precisionlymphonet,chen2024can}. CNNs have shown promising ability to perform tasks such as tumor detection, pathological classification, disease screening, and even the planning of surgery. In such cases, under ideal conditions, it can surpass human expert performance.

However, despite their numerous strengths, CNNs also have some notable drawbacks. Unlike feature-engineered models that are structured to make predictions from hand-crafted features, CNNs' extraction of features is fully automated. This fully integrated learning ability is useful in domains such as biomedical imaging, which involve complex data-driven feature selection ~\citep{litjens2017survey,anwar2018medical,chen2017deep}.

However, because CNNs are practically not interpretable, the common means of performance evaluation is through empirical experiments. These evaluation practices are commonly used in the field of machine learning, and include performance indicators such as the classification accuracy, sensitivity, specificity, F-1, confusion matrix, and more. 

However, the widespread development and adoption of CNNs in fields such as healthcare requires strict scrutiny beyond mere performance metrics. Therefore, the experimental design used in the field of machine learning might not be fully suitable for cancer pathology.

CNN architectures can exhibit biases that may lead to inaccurate clinical and research results, and can mislead even experienced researchers~\citep{obermeyer2019dissecting,dhar2021evaluation,ball2023ai}. For instance, unnoticeable changes in factors such as the position of the scanner or patient, the temperature of the charged-coupled device sensor used for the collection of images, and even the different technicians who acquire the data can lead to certain differences that are difficult to detect with the naked eye but can have a significant influence on CNNs~\citep{ball2023ai}.

In this study we examine the presence and impact of dataset bias on CNN analysis applied to cancer pathology. This is achieved by testing the ability of CNN models to identify cancer based purely on extracted background regions of original cancer pathology images that do not contain medically relevant information. The ability of a CNN to identify cancer from these background sections indicates that their performance demonstrated might be driven by bias rather than a true ability to identify cancer.

This study uses 13 well-known cancer pathology benchmark datasets that have already been cited thousands of times in the cancer pathology literature. %The analysis examines all of these datasets, not merely those that were identified to be biased, to determine the prevalence and the magnitude of CNN bias in cancer pathology. 
The study examines these datasets in combination with some of the most common CNN architectures to profile the differences between these architectures in the context of dataset bias.

%We perform this by cropping small patches from each original image in the dataset, creating five new datasets. Each of these five datasets is made of similar cropped regions across the original dataset, consisting of images without any diagnostic or anatomically relevant content. 

%We further use four common CNN models to train the resulting new datasets, and then we compare the classification performance of the original dataset images with that of the new datasets containing only background images without any biomedical information.

The analysis reveals a concerning trend, which further underscores the significance of understanding what CNN models are learning in biomedical image analysis. The classification accuracies of CNNs trained on cropped background images are noticeably higher than mere chance accuracy and are close to the classification accuracy of the trained original images in some cases. % This implies that CNNs may be using image-specific biases contained in the cropped background images to signify class labels. 

These findings show that traditional machine learning performance evaluation practices might not be ideal for the field of cancer pathology. Testing CNNs using some of the common benchmark datasets might lead to biased results that do not reflect the true ability of CNNs to identify cancer. These findings should be used as a recommendation to conduct empirical evaluations of CNN with caution. %please check intended meaning has been retained

\subsection*{Related Work}
\label{related_work}

Dataset %MDPI: If there is only one subsection within a section, it should not be numbered. We have thus removed this section number. Please confirm.  Author response: We confirm.
 bias has been discussed in the context of machine learning before deep neural networks were introduced~\citep{torralba2011unbiased}. Unbalanced datasets have been a known source of dataset bias, and several approaches have been proposed to handle unbalanced data~\citep{ganganwar2012overview}. Other biases, driven by the annotation of the dataset or the dataset design, were also noted before CNNs became common~\citep{shamir2008evaluation,torralba2011unbiased}.

As CNNs became the primary solution to image analysis, other biases driven by the method of acquisition of the images were reported~\citep{erukude2024identifying,dhar2021evaluation,dhar2022systematic}. These biases showed that CNNs are sensitive to information that cannot be sensed by the unaided human eye, yet allow CNNs to provide good classification accuracy. 

Biases were also observed in biomedical images~\citep{shamir2011assessing,obermeyer2019dissecting,dhar2021evaluation,ball2023ai}. Studies have shown that CNN models often learn to exploit superficial correlations or background artifacts in training data, learning features and patterns that are highly predictive but not directly associated with the underlying disease pathology~\citep{ball2023ai,degrave2021ai,zech2018variable}. For example, CNNs rely on hospital-origin metadata tags and radiographic positioning rather than lung pathology, as shown when trained on chest X-rays~\citep{zech2018variable}. 

This phenomenon, known as {\it shortcut learning %MDPI: Please confirm if the italics is unnecessary and can be removed.    Author response:  We would rather keep the italic because it is a term, but it is not critical. If you want to remove it that will be okay too.
}, refers to the situation when machine learning models do not learn from meaningful features to make decisions, but instead focus on easy-to-learn features that can lead to good performance on training datasets but poor generalization on unseen data. However, this has not been profiled in the context of cancer pathology via a thorough analysis of different datasets and CNN architectures.

%\section{Methodology And Experimental Design}
\section{Methods}
\label{methods}

We %MDPI: This is a reminder: Please notice that for companies/manufacturers of chemicals & reagents, devices, instruments, commercial cell lines/samples/ materials used, please add the information of “manufacture, city, abbreviated state (for USA/Canada), country”; for all the software used, please add the information of “version, manufacture, city, abbreviated state (for USA/Canada), country”.    Author response:  We did not use any company or manufacturer. No change is needed here.
 performed a series of experiments on thirteen highly used biomedical image datasets. The experiments focused on the automatic classification of radiographs using four commonly used CNN models: ResNet50~\citep{behar2022resnet50}, DenseNet121~\citep{pattanaik2022breast}, Inception V3~\citep{al2021automatic}, and VGG16~\citep{manasa2021skin}.
These architectures were selected because they are among the most commonly used CNN architectures. While they are used extensively in all domains, these architectures are widely used for the purpose of cancer pathology~\citep{ccakmak2025deep,desai2025multi,emegano2025histopathology,jusman2025comparison}.

\subsection{Experimental Design}
\label{Design}

In many cases, a radiograph has some clinically relevant parts, while other parts of the radiograph do not contain information that can allow for diagnosis. The presence of some regions that do not contain clinical information allow for the reliability of CNNs to be tested when applied to cancer pathology. If the performance evaluation is undertaken using only the parts of the radiographs that are not clinically relevant, it can provide an indication of bias in the~analysis.

Naturally, parts of the radiograph that do not contain parts of the body or other relevant medical information are not expected to allow for the diagnosis of cancer. Therefore, if the standard performance evaluation provides satisfactory results when using only medically irrelevant parts of the radiographs, that can indicate that the evaluation might be unsound. Such deceiving performance evaluation can be driven by biases that are very difficult to identify. 

To test for such biases, we compare the performance achieved with the original images to the performance observed with small sub-images that do not contain medical information. These cropped images are 20 $\times$ 20 pixel sub-images~\citep{dhar2021evaluation,dhar2022systematic} from five different parts of each original image: upper-left, upper-right, center, bottom-left, and bottom-right. That is, from each dataset, we create five new datasets with the same number of classes and the same number of images. Instead of the original images, the dataset contains small images separated from the non-medical parts of the original images. In this study, the cropping of the images was achieved with the Python Imaging Library (PIL) 11.2.1 %MDPI: Please state the version number of the software. Author response:  The version of the library has been added.
, which is a common library for basic image processing~\citep{siahaan2021implementasi}. All cropping was done in a fully automatic manner to avoid any possible human bias.

\begin{figure}[H]
%\centering
\hspace{-5pt}\includegraphics[scale=0.35]{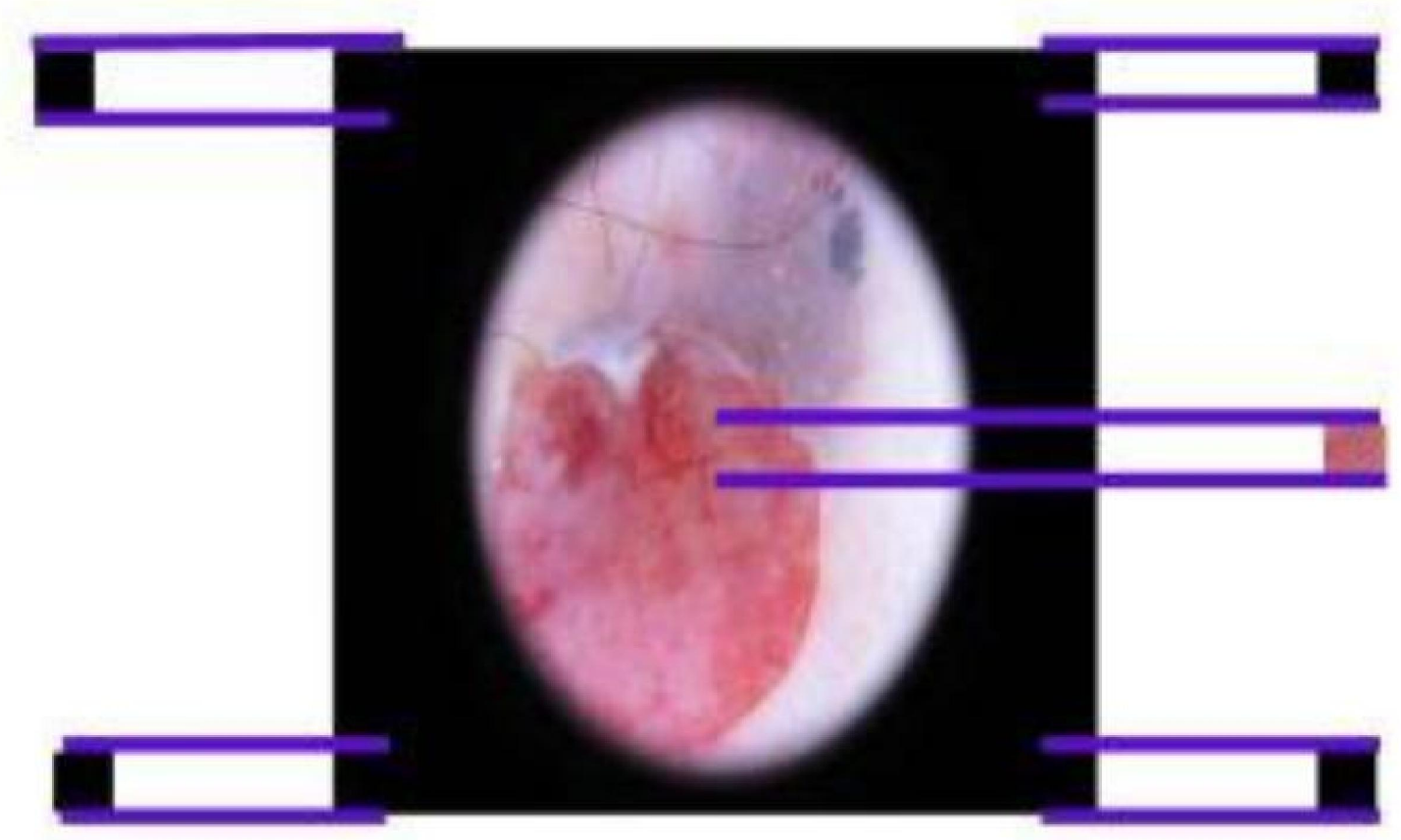}
\caption{Example %MDPI: The contents of this figure are not legible. Please replace the image with one of a sufficiently high resolution (min. 1000 pixels width/height, or a resolution of 300 dpi or higher).   Author response: We increased the resolution of the image and replaced the old image with the higher resolution image. The content of the image is taken from a widely used benchmark and that is the quality of the entire benchmark so it is difficult to improve the quality of the image dramatically, unlike the other figures that we were able to improve substantially. 
 of 20 $\times$ 20 cropped images from the ISIC 2017 Skin cancer diagnosis dataset and five sections of the original image: upper-left, upper-right, center, bottom-left, and bottom-right. These small sub-image contain no medical information, and therefore are not expected to be useful for identifying cancer to a level beyond mere chance. The empirical classification accuracy of each section is compared with that of the classification accuracy observed when using the original image.}
\label{image_separation}
\end{figure}

\begin{figure}[H]
\begin{adjustwidth}{-\extralength}{0cm}
\centering %% If there is a figure in wide page, please release command \centering
\includegraphics[scale=0.37]{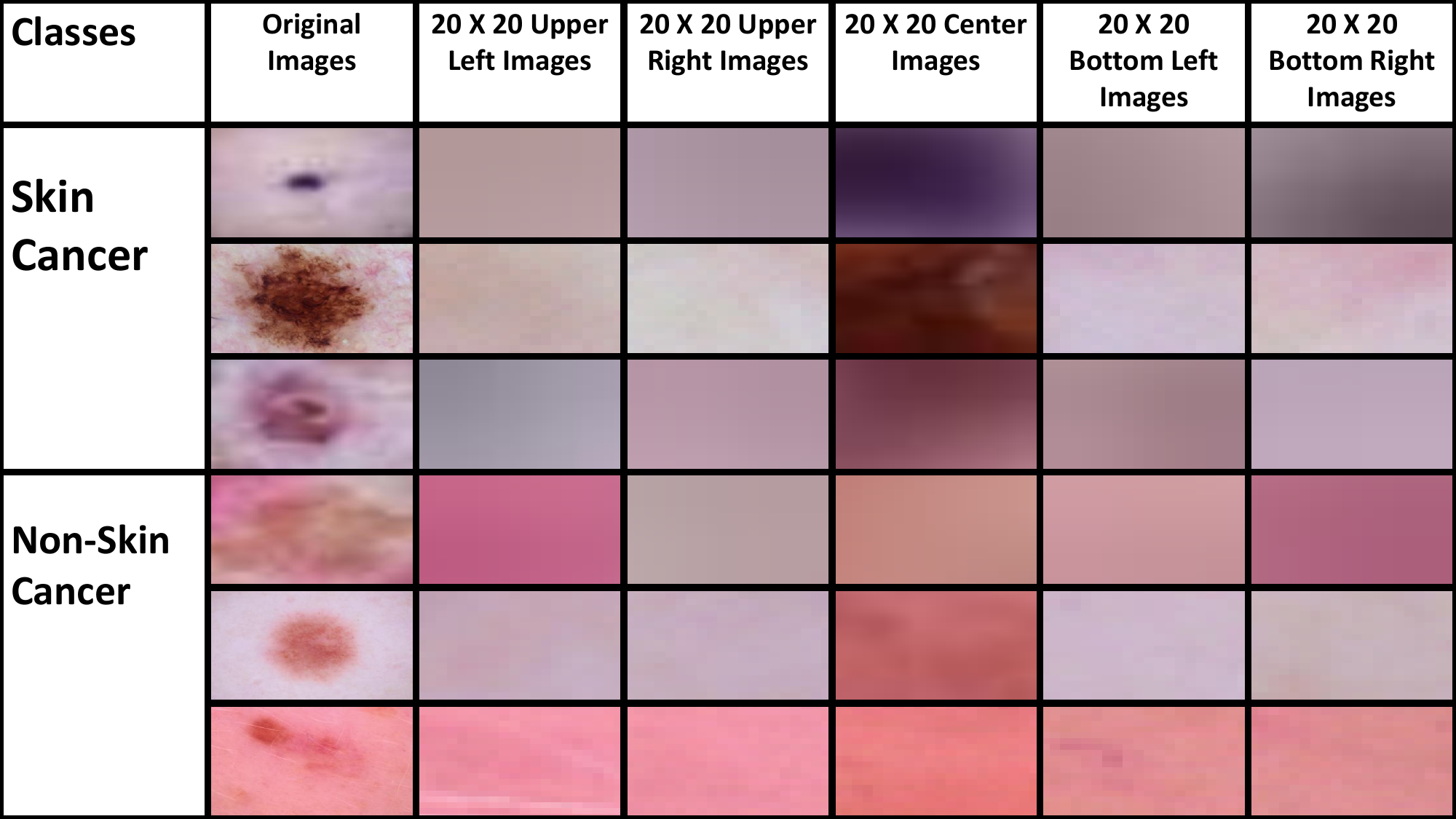}
\end{adjustwidth}
\caption{Sample %MDPI: The contents of this figure are not legible. Please replace the image with one of a sufficiently high resolution (min. 1000 pixels width/height, or a resolution of 300 dpi or higher).   Author response: We replaced the figure with a figure with a far higher resolution. The labels are now in vector graphics and should not pixelate at any zoom level. 
 images from the different classes of the DermaMNIST dataset and their respective 20 $\times$ 20 cropped sections that have no meaningful medical content. Since they do not have any medically relevant information, these non-informative sub-images are not expected to be able to identify cancer.}
\label{dataset_separation}
\end{figure}

Figure~\ref{image_separation} shows how and where each of the five 20 $\times$ 20 sub-images were cropped from the original image. Figure~\ref{dataset_separation} shows examples of the original image, with corresponding cropped five sections that depict both benign and malignant tissues, forming the basis for two classes. While the full original image allows for diagnostics, the five new datasets created from it are too small to analyze the structure of the tissue, and are not sufficiently informative to provide a reliable diagnostics.

\vspace{-6pt}

To test the ability of the CNNs to classify the images, we first split the original dataset into training and test/validation datasets using the standard 80/10/10 practice. The dataset is then tested by using the neural network architecture to evaluate the ability of the neural network to identify between positive and negative cases. The same process is repeated with each of the five datasets generated from the original dataset. The ability of the neural network to identify cancer from each of these datasets is compared to the performance when using the original dataset with the full images.

\newpage

\subsection{Model Architecture}
\label{Models}

%%%%%%These experiments are performed with several different neural network architectures. The CNN architectures were chosen for being among the most established and commonly used neural network architectures. These include ResNet-50~\citep{he2016deep}, DenseNet-121~\citep{huang2017densely,chhabra2022smart}, Inception V3~\citep{szegedy2016rethinking,mujahid2022pneumonia}, and VGG-16~\citep{Simonyan15,kaur2019automated}. 

%%%To evaluate the outcome of our experiments on the use of deep learning models for biomedical image analysis to unmask their bias, a consistent experimental design (dataset setup and training parameters) was employed across all models used in every experiment. Four different CNN architectures were selected due to their widespread application, particularly in biomedical image analysis and computer vision benchmarks. Furthermore, due to the achievements of these architectures—ResNet50 \cite{hossain2022transfer}, DenseNet121 \cite{chhabra2022smart}, Inception V3 \cite{mujahid2022pneumonia}, and VGG16 \cite{kaur2019automated}—in the field of clinical analysis. 

\textls[-10]{Unless stated otherwise, the training was performed with the commonly used practice of transfer learning using pre-trained ImageNet weights.  The input image resolution is model-dependent, and the images were adjusted to the model input. A standardized input resolution of 224 $\times$ 224 was used for ResNet-50, DenseNet-121, and VGG16, while 299 $\times$ 299 was used for Inception V3 in line with the training convention of each model.}

Modifications were made by replacing the final classification layer of each model with a fully connected layer, producing two logit models to support a two-class classification task (cancer present or cancer absent). A softmax layer was applied as the final layer to produce a probability distribution that facilitates decision-making.

All four models were trained using the standard Adam optimizer~\citep{kingma2014adam}. A constant learning rate of 0.0001 was used for each model across all experiments. This can help to achieve better stability and convergence, further improving generalization accuracy~\citep{wilson2001need}. Unless stated otherwise, the training was conducted over five epochs, which is sufficient for convergence and the use of transfer learning models. Furthermore, a batch size of 32 training samples was used in one iteration during training, balancing the models' performance between computational efficiency and quality of learning. 

As discussed above, each dataset provided five additional datasets, such that each new dataset was made from only cropped images containing a certain part of the original images. For consistency, the exact same practices were used for each of these new datasets, including the cropping, resolution, and architecture parameter settings as described above. The same parameters were also used for testing the original images.

The computing was carried out using the Beocat computing facility~\citep{hutson2019managing}. Beocat is a powerful cluster with over 11,000 CPU cores. To train the CNNs, we used GPUs. Beocat has 90~GPU-enabled nodes that host 170 GPUs, with models ranging from the simple GeForce GTX 1080 Ti to the advanced nVidia L40S. This cluster provides powerful computing that can enable a large number of CNN experiments.

\subsection{Biomedical Images Datasets}
\label{data}

We obtained thirteen publicly available benchmark biomedical image datasets frequently used for developing and testing CNN methodology for cancer diagnosis. In these experiments, we use these datasets to evaluate the performance of CNN architectures on cancerous versus noncancerous tissues. 

The datasets were selected based on their prevalence as benchmarks for developing and testing CNN-based methods for cancer pathology. The datasets that were chosen were all highly cited in the literature, and therefore an analysis of these datasets is required to ensure that they provide a sound platform for the development and testing of reliable~CNNs.

These datasets also represent a diversity of types of cancers, modalities, magnifications, and more. Table~\ref{tab: MedMNIST} shows the different types of cancers included in the experiments, including colorectal tissues, lung tumors, and skin and breast cancer. Modalities include microscopy Hematoxylin and Eosin (H\&E) such as in BreakHis, as well as Computed Tomography (CT), X-ray, and ultrasound, as included in the MedMNIST datasets, or the digital camera dermoscopic images included in the ISIC datasets. These cover a broad range of modalities, which are the common modalities used for cancer pathology.

\begin{table*}[h]
\centering
\tiny
\renewcommand{\arraystretch}{1.5}
%\resizebox{\textwidth}{!}{
%{\bfseries
% \begin{tabular}{|c|c|c|p{6cm}|c|c|c|}
\begin{tabular}{|c|c|c|l|c|c|c|}

\hline
%\multicolumn{7}{|c|}{\textbf{Cancer MedMNIST Dataset}} \\
\hline
\rowcolor[gray]{0.9}
% \textbf{Dataset} & \textbf{Data Modality} & \textbf{Initial Class\Label} & \textbf{Modified Binary Class\Label} & \textbf{Total Samples} & \textbf{Cancer Present Samples} & \textbf{Cancer Absent Samples} \\
\hline
 PathMNIST& Histopatology & Multi-Class (9) & 1. Cancer-Associated Stroma, 2. Colorectal Adenocarcinoma Epithelium = Cancer Present / 3. Adipose Tissue, 4. Background, 5. Debris, 6. Lymphocytes, 7. Mucus, 8. Smooth Muscle and Normal Colon Mucosa = Cancer Absent & 107,180 & 95,435 & 11,745\\
\hline
DermaMNIST & Dermatoscope & Multi-Class (7) & 1. Melanoma 2. Basal cell carcinoma = Cancer Present / 3. Melanocytic nevi, 4. Benign keratosis-like lesions, 5. Dermatofibroma, 6. Vascular lesions 7. Actinic keratoses = Cancer Absent  & 10,015 & 656 & 9,359\\
\hline
BreastMNIST& Breast Ultrasound  & Multi-Class (3)  & 1. Malignant = Cancer Present / 2. Normal, 3. Benign = Cancer Absent & 780 & 570 & 210\\
\hline
NoduleMNIST3D& Chest CT & Binary-Class (2) & 1. Malignant = Cancer Present / 2. Benign = Cancer Absent & 1,633 & 401  & 1232\\
\hline
\end{tabular}
%}
%}
\caption{The description, class modification, and distribution of samples of the MedMNIST Dataset.}
\label{tab: MedMNIST}
\end{table*}

\subsection{MedMNIST Datasets}
\label{MNIST}

The MedMNIST+ dataset~\citep{medmnistv2} is a larger version of the MedMNIST dataset in terms of image resolution. This improves on the 28 $\times$ 28 and 28 $\times$ 28 $\times$ 28 of the MedMNIST dataset~\citep{medmnistv2,doerrich2025rethinking}. MedMNIST is a standardized collection of eighteen biomedical image datasets for classification with various settings, diversifying the evaluation across pixel size, modalities, and classification tasks~\citep{medmnistv2}. 

Four datasets from MedMNIST+ were used. These are pathological images suitable for the classification of positive or negative cancer cases. They include the PathMNIST~\citep{kather2019predicting}, DermaMNIST~\citep{tschandl2018ham10000}, BreastMNIST~\citep{al2020dataset}, and NoudleMNIST~\citep{samuel2011lung} datasets. Table~\ref{tab: MedMNIST} shows the structure of these datasets and their respective modification into a binary class. These datasets can be publicly accessed at %MDPI: Please add the access date (format: Date Month Year), e.g., accessed on 1 January 2020. The same for below.  Author response: access date has been added.
 \url{https://zenodo.org/records/10519652}.

\subsection{BreakHis Datasets}
\label{BreakHis}

BreakHis, the Breast Cancer Histopathological Image Classification, is publicly available at \url{https://www.kaggle.com/datasets/ambarish/breakhis}. It is a breast cancer histopathology dataset widely used for machine learning research, specifically for classifying images of breast cancer tissues. These histopathology images were collected from 82 patients, containing 7909 image samples~\citep{spanhol2015dataset}. The BreakHis dataset contains different magnification levels: $40\times$, $100\times$, $200\times$, and $400\times$. Each magnification level consists of eight classes, which we modified to suit a binary classification task as described in Table~\ref{tab:BreakHis}.

\begin{table}[H]
\caption{The %MDPI: Please add an explanation for background color in the table footer. If the background color is unnecessary, please remove it.   Author response: color has been removed.
 Description, class modification, and distribution of samples of the BreakHis Dataset. Figures assigned to each class are as follows: Adenosis = 1; Fibroadenoma = 2; Phyllodes tumor = 3; Tubular adenoma = 4; Ducta carcinoma = 5; Lobular carcinoma = 6; Mucinous carcinoma = 7; and Papillary carcinoma = 8.}
\label{tab:BreakHis}
\footnotesize

\renewcommand{\arraystretch}{1.8}

\begin{adjustwidth}{-\extralength}{0cm}
\centering %% If there is a figure in wide page, please release command \centering

%\begin{tabularx}{\fulllength}{Cp{2cm}cp{4cm}CCC}
\begin{tabular}{ccccccc}

\toprule
\multicolumn{7}{c}{\vspace{2pt}\textbf{BreakHis Dataset}} \\
\hline
%\rowcolor[gray]{0.9}
\textbf{Magnification} & \textbf{Modality} & \textbf{Initial Class} & \textbf{Modified Binary Class} & \textbf{Total Samples} & \textbf{Cancer Present Samples} & \textbf{Cancer Absent Samples} \\
\hline
$40\times$  & \multirow{4}{*}{\makecell[l]{Microscopy\\Histopathology}} & \multirow{4}{*}{Multi-Class (8)} & \multirow{4}{*}{\makecell[l]{1 to 4 = Cancer Absent/\\5 to 8 = Cancer Present}} & 1995  & 1370 & 652 \\
\cline{1-1} \cline{5-7}
$100\times$ &                      &                       &                    & 2081  & 1437 & 644 \\
\cline{1-1} \cline{5-7}
$200\times$  &                       &                       &                    & 2013 & 1390 & 623   \\
\cline{1-1} \cline{5-7}
$400\times$ &                       &                       &                      & 1820  & 1232 & 588 \\
\bottomrule
%\end{tabularx}
\end{tabular}
\end{adjustwidth}
\end{table}

\newpage
\subsection{International Skin Imaging Collaboration Datasets}
\label{ISIC}

The International Skin Imaging Collaboration (ISIC) is a benchmark dataset associated with a yearly competition for machine learning experts.  It consists of a yearly release of dermatoscopic images from 2016 to 2020. This provide standardization for skin imaging, creating an open-access repository of biomedical images.  We obtained four datasets, ISIC-2016~\citep{gutman2016skin}, ISIC-2017~\citep{codella2018skin}, ISIC-2018~\citep{tschandl2018ham10000,codella2018skin}, and ISIC-2019~\citep{tschandl2018ham10000,codella2018skin,combalia2019bcn20000}, for evaluation, as shown in Table~\ref{tab: ISIC}. These datasets are publicly available at \url{https://challenge.isic-archive.com/}.

\begin{table}[H]
\footnotesize
\caption{The %MDPI: Please add an explanation for background color in the table footer. If the background color is unnecessary, please remove it.    Author response: color has been removed.
 description, class modification, and distribution of samples of the ISIC dataset.}
\label{tab: ISIC}
\renewcommand{\arraystretch}{1.5}

\begin{adjustwidth}{-\extralength}{0cm}
\centering %% If there is a figure in wide page, please release command \centering

%\begin{tabularx}{\fulllength}{Cccp{6cm}CCC}
\begin{tabular}{ccccccc} 
\toprule
\multicolumn{7}{c}{\vspace{3pt}\textbf{The ISIC (2016--2019) Dataset }} \\
\hline
%\rowcolor[gray]{0.9}
\textbf{Dataset} & \textbf{Data Modality} & \textbf{Initial Class} & \textbf{Modified Binary Class} & \textbf{Total Samples} & \textbf{Cancer Present Samples} & \textbf{Cancer Absent Samples} \\
\hline
 ISIC-2016 & Dermatoscope & Binary Class & 1. Melanoma = Cancer Present; 
 2. Benign = Cancer Absent & 1279 & 248 & 1031\\
\midrule
 ISIC-2017 & Dermatoscope & Multi-Class (3) & 1. Melanoma = Cancer Present; 
 2. Seborrheic\_keratosis;  3. Nevus = Cancer Absent & 11,527 & 3736 & 7790\\
\midrule
ISIC-2018 & Dermatoscope & Multi-Class (7) & 1. Melanoma; 2. Basal cell carcinoma = Cancer Present; 3. Melanocytic nevi; 4. Benign keratosis-like lesions; 5. Dermatofibroma; 6. Vascular lesions; 7. Actinic keratoses = Cancer Absent  & 10,015 & 7007 & 1003\\
\midrule
ISIC-2019 & Dermatoscope  & Multi-Class (8)  & 1. Malanoma; 2. Besal Cell Carcinoma; 3. Antinic Keratosis; 4. Squamous Cell Carcinoma = Cancer Present; 5. Melanocytic Nevus; 6. Benign Keratosis; 7. Dermotafibroma; 8. Vascular Lesion = Cancer Absent & 780 & 546 & 78\\
\bottomrule
%\end{tabularx}
\end{tabular}
\end{adjustwidth}

\end{table}

\subsection{Breast Histopathology Images for Invasive Ductal Carcinoma (IDC) Dataset}
\label{IDC}

The Breast Histopathology Image (IDC)~\citep{abdallah2023enhancing} dataset is a benchmark for Invasive Ductal Carcinoma (IDC) detection, specifically designed to classify breast histopathological images as either cancerous (IDC-positive) or non-cancerous (IDC-negative), as shown in Table~\ref{tab: IDC}. The dataset comprises 162 whole-slide images (WSIs) obtained from breast cancer patients, totaling 277,524 images~\citep{Janowczyk2016DeepLearningDP}
. The dataset is publicly available at \url{https://www.kaggle.com/datasets/paultimothymooney/breast-histopathology-images}.

\begin{table}[H]
\small
\caption{Distribution %MDPI: Please add an explanation for background color in the table footer. If the background color is unnecessary, please remove it. Author response: Color has been removed.
 of samples of the Breast Histopathology Image (IDC) Dataset.}
\label{tab: IDC}
\renewcommand{\arraystretch}{1.5}

\begin{adjustwidth}{-\extralength}{0cm}
\centering %% If there is a figure in wide page, please release command \centering

%\begin{tabularx}{\fulllength}{CCCCCCC}
\begin{tabular}{ccccccc}

\toprule
\multicolumn{7}{c}{\vspace{3pt}\textbf{ Breast Histopathology Image (IDC) Dataset}} \\
\hline
%\rowcolor[gray]{0.9}
\textbf{Dataset} & \textbf{Modality} & \textbf{Initial Class} & \textbf{Modified Binary Class} & \textbf{Total Samples} & \textbf{Cancer Present Samples} & \textbf{Cancer Absent Samples} \\
\hline
Breast Histopathology Image (IDC)  & {Microscopy Histopathology} & Binary  & IDC-Nagative = Cancer Absent/IDC-Positive = Cancer Present & 277,524 & 78,786 & 198,738 \\
\bottomrule
%\end{tabularx}
\end{tabular}
\end{adjustwidth}

\end{table}

\newpage
\section{Results}
\label{results}

\subsection{CNNs and MedMNIST+ Datasets}
\label{MedMNIST Results}

The experiments evaluated the four CNN architectures (ResNet50, DenseNet121, InceptionV3, and VGG16) across four MedMNIST+ datasets. Each MedMNIST+ dataset represents a different type of cancer and imaging modality. 

% The experiments were first performed on the full original images. The performance observed using the original images served as baseline performance. We then experimented on the cropped image datasets. The comparison between the performance observed with the original images and the performance observed with the cropped datasets was use to test whether classification accuracy depends mostly on disease-specific features or background characteristics/imaging artifacts.  

\begin{figure}[H]
\includegraphics[scale=0.2]{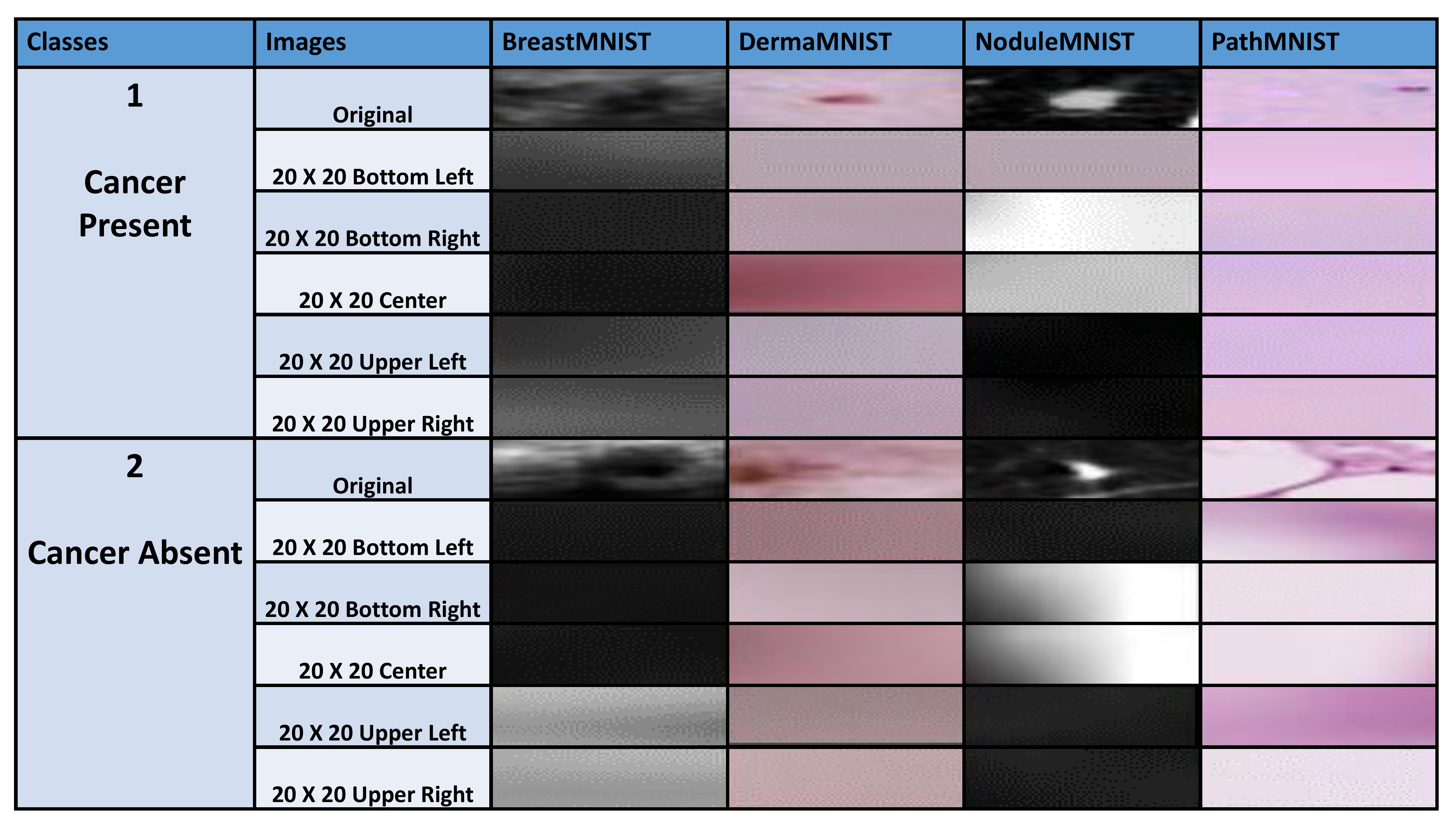}
\caption {Sample %MDPI: (1) The contents of this figure are not legible. Please replace the image with one of a sufficiently high resolution (min. 1000 pixels width/height, or a resolution of 300 dpi or higher). (2) We moved the Figures 5--7 here after the Figure 4, please revise their citation in numerical order.    Author response: 1. We replaced the figure with another figure with a much higher resolution. The labels are in vector graphics so thy look clear at any resolution. 2. We saw the problem with the ordering of the figures, but the figure should be in the right place. 
 images across the MedMNIST datasets, showing the original images and their \mbox{20 $\times$ 20} cropped images. Mostly, the cropped images contain little or no medical content useful for correct diagnostics.}
\label{Figure MedMNIST_Samples}
\end{figure}

The cropped-image datasets consisted of 20 $\times$ 20 pixel extractions from five different sections of the original images (bottom-left, bottom-right, center, upper-left, and upper-right). Due to the differences in lesion size, shape, and location, they often include only background or imaging artifacts, as shown in Figure%MDPI: Figure citation should be in order. Figure shoudl be after figure 4. please modify.   Author response: The figure has been moved right below this text. The order oof the figure is now correct.
~\ref{Figure MedMNIST_Samples}.

\vspace{-7pt}

\subsubsection{BreastMNIST (Ultrasound Scan Imaging)}
\label{Result_BreastMNIST}

The BreastMNIST~\citep{al2020dataset} dataset contains two classes: benign and malignant breast tumors. The images are in grayscale echotextures consisting of tissue layers, shadow artifacts, and boundaries. Figure~\ref{Figure MedMNIST_Samples} shows sample images, comparing the original and the 20 $\times$ 20 cropped-image datasets. Each cropped-image dataset consists of black or gray background images. These background images do not contain tumor structures that would be visible to an expert radiologist to allow for meaningful diagnostics. We do not expect deep neural networks to identify positive and negative cases from these cropped images. 

\begin{figure}[H]

\hspace{-18pt}\includegraphics[width=0.9\linewidth]{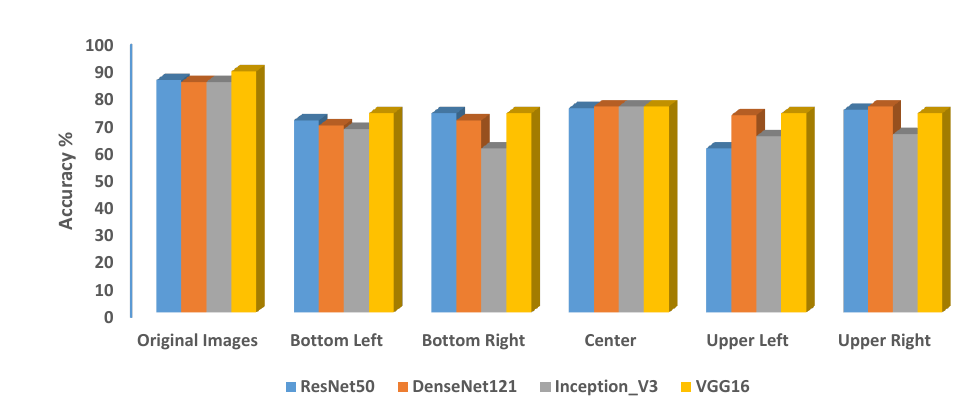}
\caption {The %MDPI: Please check all figures and ensure that there are no duplicated figures.  Author response: We were not able to identify identical images.
 classification accuracy of the BreastMNIST original image dataset alongside cropped 20 $\times$ 20 pixel sections. As expected, CNNs models can classify the original images. Surprisingly, all CNN models performed better than the random chance accuracy of 50\%, when applied to the cropped sections, highlighting that these models can classify images that lack lesion-specific features.}
\label{Figure_BreastMNIST}
\end{figure}

The results from the original images show that CNN architectures can achieve a classification accuracy as high as $\sim$88.46\%. To test for bias, we also experimented on the \mbox{20 $\times$ 20 pixel} image datasets. The results revealed that all CNN models can classify the cropped images with an accuracy well above the mere chance of $50\%$. This performance for a binary classification is not expected, because the cropped images often exclude the tumor region. For example, Figure~\ref{Figure_BreastMNIST} shows that CNNs can identify cancerous and noncancerous tumors from background images with an accuracy as high as $\sim$75.64\%. That shows that cancer can be identified with an accuracy better than mere chance even when no clinically relevant information is included in the data. Interestingly, VGG16 achieved a consistent performance of $\sim$73.08\% across all cropped-image datasets except the center images. % The result also reveals that CNN models achieved a significantly higher performance on the cropped center images compared to the corner image datasets (bottom-left, bottom-right, upper-left, and upper-right). 

\vspace{-9pt}

Table~\ref{BreastMNIST_balanced} shows the classification accuracy, precision, recall, and F-1 scores of the BreastMNIST dataset when the dataset is balanced. The observation that the CNN is able to identify cancer with an accuracy higher than mere chance indicates that information in these images that is not medical or related to cancer can be used by the CNN to make a detection. That shows that the detection of cancer using the dataset might not necessarily prove that the CNN can indeed detect cancer, but that it uses other information in these images that is not related to the presence of cancer.

\begin{table}[H]
\small
\caption{Accuracy, precision, recall, and F-1 results of the analysis of the balanced BreastMNIST.}
\label{BreastMNIST_balanced}

%\renewcommand{\arraystretch}{2.5}

%\begin{adjustwidth}{-\extralength}{0cm}
%\centering %% If there is a figure in wide page, please release command \centering
%\begin{tabularx}{\textwidth}{lcCCCC}
\begin{tabular}{lccccc}

\toprule
\textbf{Model %MDPI: We added bold for the header, please confirm.  Author response: We confirm.
} & \textbf{Sub Dataset }& \textbf{Accuracy} & \textbf{Recall} & \textbf{F}1 & \textbf{Precision} \\
\midrule
DENSENET121 & Bottom Left & 85.8372 & 85.5634 & 81.9552 & 78.7535\\
DENSENET121 & Bottom Right & 62.8590 & 4.4366 & 7.4707 & 28.3412\\
DENSENET121 & Center & 62.7399 & 3.9437 & 6.8230 & 35.0489\\
DENSENET121 & Original Images & 61.2442 & 6.5141 & 10.0572 & 64.2105\\
DENSENET121 & Upper Left & 62.9782 & 4.6479 & 7.9069 & 57.9132\\
DENSENET121 & Upper Right & 62.8855 & 4.7887 & 7.9559 & 42.9699\\
INCEPTION\_V3 %MDPI: Please confirm the V3 format, please check the full text.    Author response: We confirm..
 & Bottom Left & 84.8312 & 84.7887 & 80.8230 & 77.6369\\
INCEPTION\_V3 & Bottom Right & 62.6208 & 2.8169 & 4.5410 & 41.0638\\
INCEPTION\_V3 & Center & 62.7796 & 2.6761 & 4.5365 & 24.9397\\
INCEPTION\_V3 & Original Images & 62.2766 & 2.3592 & 4.1722 & 19.4480\\
INCEPTION\_V3 & Upper Left & 62.7929 & 2.5000 & 4.2595 & 29.0031\\
INCEPTION\_V3 & Upper Right & 62.7929 & 2.6761 & 4.5215 & 25.6140\\
RESNET50 & Bottom Left & 85.1621 & 83.2042 & 80.8063 & 78.8106\\
RESNET50 & Bottom Right & 63.0443 & 6.9014 & 11.4470 & 58.2067\\
RESNET50 & Center & 63.0841 & 5.8451 & 9.9479 & 44.7500\\
RESNET50 & Original Images & 61.7737 & 1.6197 & 2.8298 & 13.2013\\
RESNET50 & Upper Left & 63.6929 & 8.2042 & 13.4575 & 57.7125\\
RESNET50 & Upper Right & 63.6664 & 7.9930 & 13.1650 & 68.3775\\
VGG16 & Bottom Left & 82.9385 & 82.4296 & 78.4074 & 74.9589\\
VGG16 & Bottom Right & 63.0311 & 7.9930 & 11.7898 & 42.2358\\
VGG16 & Center & 63.0179 & 7.2183 & 10.9649 & 32.5476\\
VGG16 & Original Images & 62.8326 & 6.6549 & 11.3717 & 62.9896\\
VGG16 & Upper Left & 63.3091 & 7.8521 & 11.8730 & 63.5431\\
VGG16 & Upper Right & 63.2694 & 7.6056 & 11.5838 & 53.5498\\
\bottomrule

%\end{tabularx}
\end{tabular}
%\end{adjustwidth}

\end{table}

\subsubsection{DermaMNIST (Dermoscopic Skin Lesion Imaging)}
\label{DermaMNIST_Result}

The DermaMNIST~\citep{tschandl2018ham10000} dataset was collected from two different sites spanning over 20 years. Some of the images were collected before the use of digital cameras. It consists of 10015 dermoscopic images from the HAM10000, a collection of common pigmented skin lesions. As described above, we derived the cropped-image datasets from the original image dataset. The bottom-left, bottom-right, center, upper-left, and upper-right images predominantly contained skin-tone background, and did not contain structures related to the skin lesion. In contrast,  the center image dataset occasionally included small portions of the lesion. However, these tissue structures are too small to allow for meaningful diagnostics, as shown in Figure~\ref{Figure MedMNIST_Samples}.

All CNN architectures can classify the original images with very high accuracy, ranging from $\sim$94.36\% to $\sim$95.01\%. These scores are expected, and were achieved by VGG16 and DenseNet121, respectively. However, CNNs are not expected to classify the cropped-datasets, as those are mostly skin-colored background images that lack medical content. 

\begin{figure}[H]

\hspace{-18pt}\includegraphics[width=0.9\linewidth]{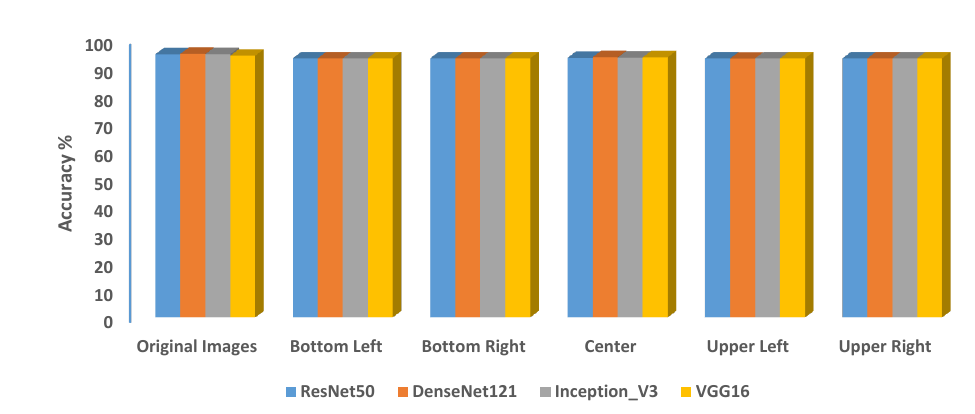}
\caption {The classification accuracy of the original image alongside a cropped 20 $\times$ 20 pixel of the DermaMNIST datasets. All CNN models achieved a consistent classification accuracy of 93.42\% on all four corner-cropped-image datasets, even when they contained only skin-toned background images. However, performance is slightly higher and varied on center-cropped images across all models; these images may contain small fractions of the lesion structure.}         
\label{Figure DermaMNIST}
\end{figure}

Surprisingly, all CNN models can also identify positive and negative cases  across the cropped-image datasets, with an accuracy very close to that achieved by the original images. The results are shown in Figure%MDPI: There is mentioned Figure 6 before Figure 4, figures should be cited in numerical order, please check and revise.   Author response: We changed the order of the figures.
~\ref{Figure DermaMNIST}. Furthermore, all CNN models achieved a consistent accuracy score of $\sim$93.42\% across the cropped-image datasets, except for the center image datasets. Here, performance varied across CNN models and was slightly higher than their performance on the corresponding cropped images datasets.   

\newpage
The results reveal that the classification accuracy is consistently biased across the four corners crops (bottom-left, bottom-right, Upper-left, and upper-right). That is, cancer can be identified even when it is clear that the images do not include any medically relevant information. This suggests that the neural networks identify the presence of the disease not merely through clinical information, but also from other pieces of information and artifacts that might be linked to the way the datasets were collected rather than the clinical condition of the patient. Therefore, the classification accuracy observed when using CNNs to identify cancer might be overoptimistic, and not a direct reflection of the CNN models to automatically identify cancer. 

% neural networks may be relying on skin-toned backgrounds, lighting gradients, or pattern distribution unique to both classes. Rather than identifying the lesion morphology, models may be detecting data acquisition-dependent features.  

In contrast to the small corner patches, the performance varied across all CNN models on the center images. This reflects the structural characteristics of the dataset, where the corners are generally uniform and homogeneous, while the center often contains lesion content. However, as shown through the corner patches, even in the absence of lesion structures in the cropped-image datasets, classification accuracy remained high.

\subsubsection{NoduleMNIST (CT Scan Imaging)}
\label{Result_NoduleMNIST}

NodueMNIST~\citep{samuel2011lung} dataset consists of thoracic CT scan slices. The images are centered on lung nodules labeled for binary classification (benign and malignant cases). The images show high contrast between lung functional tissues and surrounding tissues. If present, nodules are typically centered in the field of view. This is due to standardized radiology workflow and processing, ensuring the lesion is in view. We extracted cropped-image datasets from the original images. The resulting datasets consist of small images that are too visually subtle to contain any medical content to the unaided eye, as shown in Figure~\ref{Figure MedMNIST_Samples}.

\begin{figure}[H]

\hspace{-20pt}\includegraphics[width=0.9\linewidth]{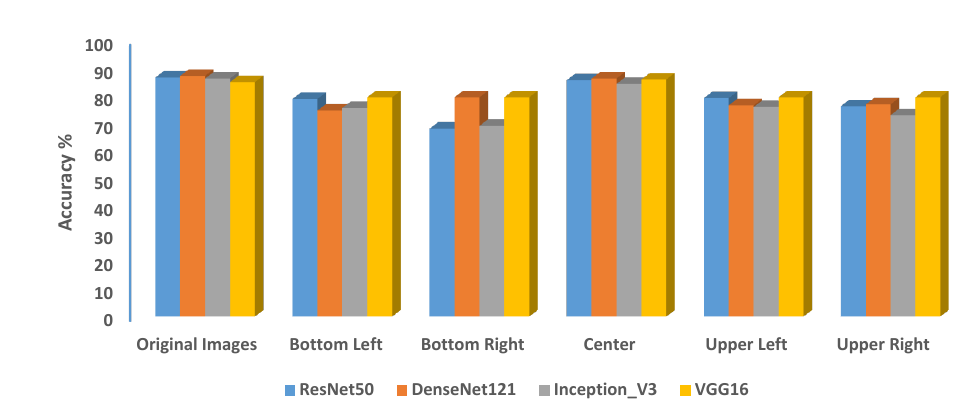}
\caption {The classification accuracy performance of CNN models on the NoduleMNIST original images and their 20 $\times$ 20 pixel subset images. Unexpectedly, neural networks can identify positive and negative cases from the cropped-image datasets, despite the lack of sufficient nodule structures to make adequate diagnostics.}         
\label{Figure NoduleMNIST}
\end{figure}

The performance across all models when using the original images is between $\sim$87.1\% and $\sim$84.84\%, achieved by DenseNet121 and VGG16, respectively. When using the five cropped-image datasets, the performance was well above mere chance, as shown in Figure~\ref{Figure NoduleMNIST}. The performance observed with the patches taken from the center of each image closely matched those of the original images. In a particular instance, it was slightly higher than the original images. For example, VGG16 achieved a performance of $\sim$84.84\% and $\sim$85.81\% on original and center images, respectively. That shows that these architectures can identify cancer without medical information.

While the high performance by CNN models on full CT images can be expected, it is surprising that CNN architectures can also accurately classify  the cropped images. %The significantly higher accuracy achieved on patches taken from the center of the images compared to patches taken from the corner of the images underscores the presence of spatial bias. This may be introduced during image acquisition, since lesions are centered in the NoduleMNIST dataset.
The classification accuracy observed when using patches taken from the center of the images is higher than the accuracy observed with patches taken from the corner of the images. This could be linked to the spatial structure of the image, where center-cropped images may contain parts of the nodule, while corner patches mostly contain only background lung tissues. These cropped-image datasets do not contain medical information that is diagnostically relevant to an expert radiologist. This suggests that deep neural networks may rely on positional cues rather than learning from pathology-specific features.

\begin{figure}[H]
\includegraphics[width=0.9\linewidth]{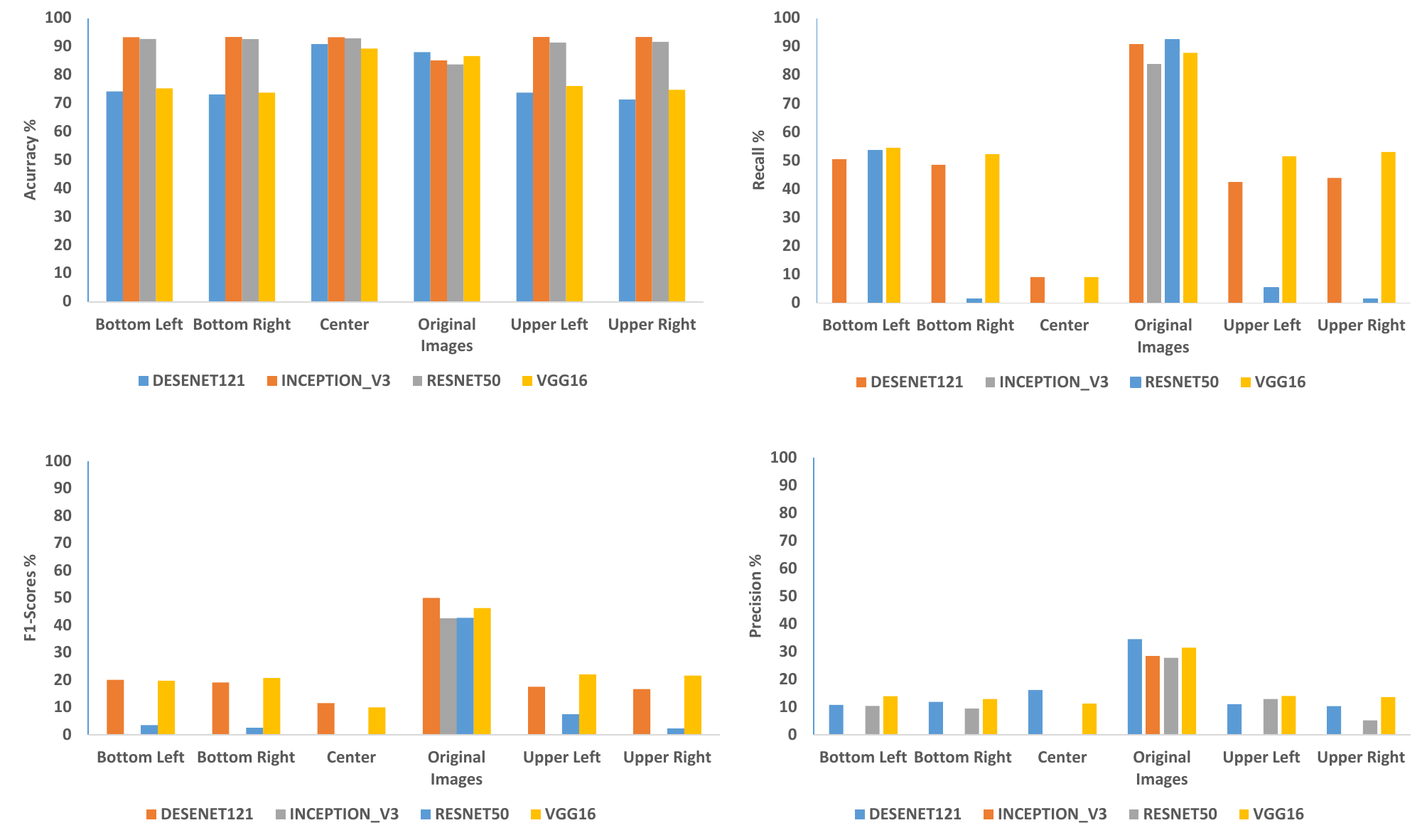}
\caption {The classification accuracy, precision, recall, and F-1 scores for the DermaMNIST dataset when each class contains 458 images.}
\label{DermaMNIST_balanced}
\end{figure}

Another experiment was performed where the number of images in each class was equal, and set to 458. Figure~\ref{DermaMNIST_balanced} shows the classification accuracy, precision, recall and F-1 %MDPI: Please unify the format should be F1 or F-1. Please check the full text.   Author response: F-1 is the correct form. We changed that throughout the paper.
 scores when each class contains exactly 458 images. The training and testing used the same settings as all other experiments. As the figure shows, the classification accuracy is far higher than mere chance. Even when using the background patches, the classification accuracy was, in many cases, higher than 90\%.

\vspace{-15pt}

\subsubsection{PathMNIST (Histopathology Slides Imaging)}
\label{Result_PathMNIST}

PathMNIST~\citep{kather2019predicting} consists of colored histopathology whole-slide images (WSIs) of colorectal cancer tissues. Each image is a single patch of tissue stained with hematoxylin and eosin (H\&E). The generation and digitization of these WSIs are achieved using a standardized microscopy workflow in clinical labs.  The images exhibit consistent staining, color balance, and pattern textures across the same class.

\begin{figure}[H]

\hspace{-20pt}\includegraphics[width=1.0\linewidth]{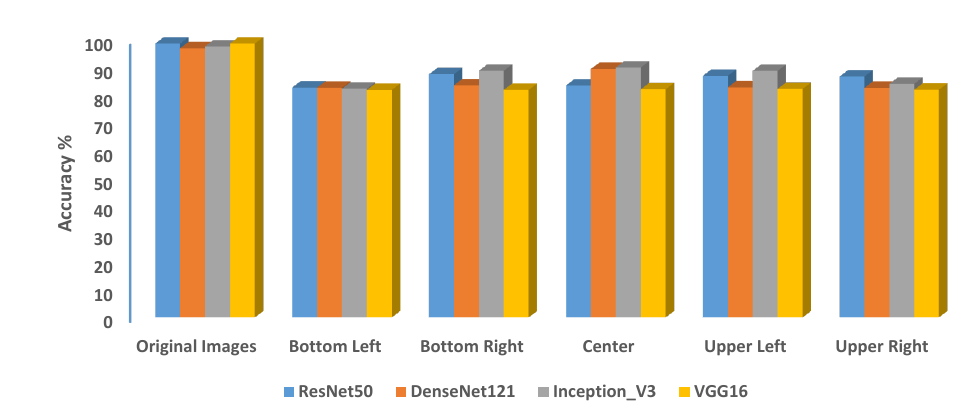}
\caption {The classification accuracy performance of CNN models on the PathMNIST original dataset and its 20 $\times$ 20 pixel subsets, suggesting that these models may be classifying histopathology images as either cancerous or noncancerous based on non-informative medical cues.}         
\label{Figure PathMNIST}
\end{figure}

The results of the experiment on the original histopathology images, as expected, show that CNNs can classify the images with an accuracy as high as $\sim$98.72\%. Figure~\ref{Figure PathMNIST} shows the results when the experiment was performed on the cropped-image datasets. InceptionV3 achieved an accuracy performance as high as $\sim$90.07\% on the center-cropped images, while VGG16 achieved the lowest performance of $\sim$81.98\% on the bottom-left cropped images. 

The results of the experiments reveal that CNN models can identify positive cases even when using small patches from the corner of the images. %This may exhibit stain-based or texture-based bias in the training of the datasets. 
A possible explanation for this is that cropped images retain color and texture features that correlate with class labels, even when they lack disease-related features. This may affect the model's performance in real-world clinical settings, where test data may differ in structure.

\vspace{-6pt}

{
Figure~\ref{PathMNIST_balanced} and Table~\ref{PathMNIST_balanced_table} show the classification accuracy, precision, recall and F-1 scores when the number of images in each class is equal, and set to 9366. The results shows that all architectures could identify cancer with an accuracy much higher than mere chance, and the classification accuracy in all cases was higher than 80\%. 

\begin{figure}[H]

\includegraphics[width=0.9\linewidth]{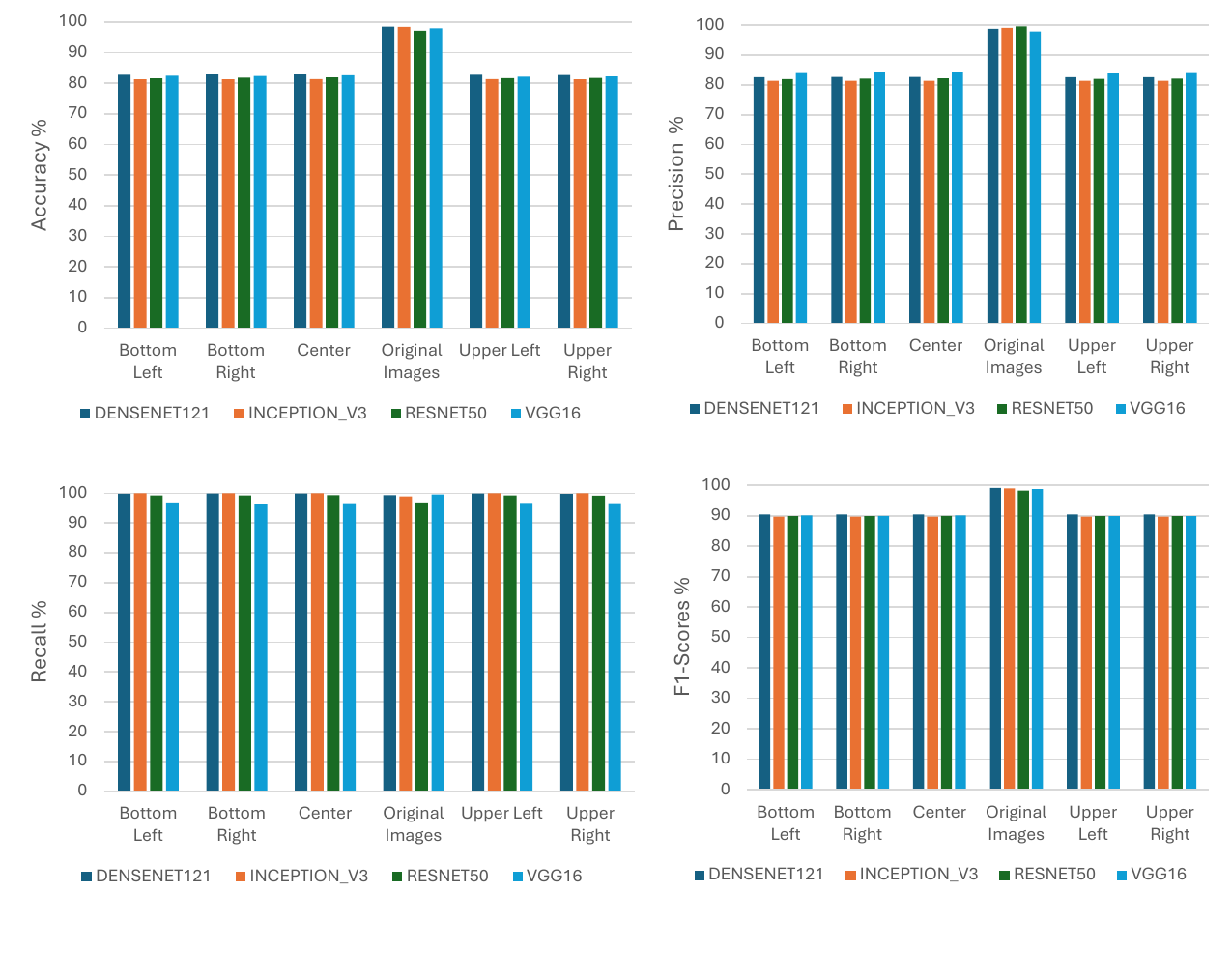}
\caption {The classification accuracy, precision, recall,  and F-1 scores for the PathMNIST dataset when the size of each class is set to 9366 images.}
\label{PathMNIST_balanced}
\end{figure}

\vspace{-6pt}
\begin{table}[H]
\caption{Accuracy, precision, recall, and F-1 results of the analysis of the balanced PathMNIST.}
\label{PathMNIST_balanced_table}
%\renewcommand{\arraystretch}{2.5}
%\begin{tabularx}{\textwidth}{lcCCCC}
\begin{tabular}{lccccc}

\toprule
\textbf{Model %MDPI: We added bold for the header, please confirm.  Author response: We confirm.
} & \textbf{Sub Dataset} & \textbf{Accuracy} & \textbf{Recall} & \textbf{F-1} & \textbf{Precision} \\
\midrule
DENSENET121 & Original Images & 98.5376 & 98.8588 & 99.3495 & 99.1036 \\
DENSENET121 & Bottom Left & 82.7994 & 82.6044 & 99.8973 & 90.4315 \\
DENSENET121 & Bottom Right & 82.8830 & 82.6653 & 99.9144 & 90.4751 \\
DENSENET121 & Center & 82.9109 & 82.6795 & 99.9315 & 90.4906 \\
DENSENET121 & Upper Left & 82.7994 & 82.5952 & 99.9144 & 90.4330 \\
DENSENET121 & Upper Right & 82.7855 & 82.6481 & 99.7946 & 90.4156 \\
INCEPTION\_V3 & Original Images & 98.4540 & 99.1425 & 98.9558 & 99.0491 \\
INCEPTION\_V3 & Bottom Left & 81.3649 & 81.3649 & 100 & 89.7251 \\
INCEPTION\_V3 & Bottom Right & 81.3649 & 81.3649 & 100 & 89.7251 \\
INCEPTION\_V3 & Center & 81.3928 & 81.3876 & 100 & 89.7389 \\
INCEPTION\_V3 & Upper Left & 81.3788 & 81.3762 & 100 & 89.7320 \\
INCEPTION\_V3 & Upper Right & 81.3649 & 81.3649 & 100 & 89.7251 \\
RESNET50 & Original Images & 97.2423 & 99.6656 & 96.9360 & 98.2818 \\
RESNET50 & Bottom Left & 81.6435 & 81.9582 & 99.2982 & 89.7988 \\
RESNET50 & Bottom Right & 81.8524 & 82.1413 & 99.2811 & 89.9016 \\
RESNET50 & Center & 81.9916 & 82.2122 & 99.3667 & 89.9791 \\
RESNET50 & Upper Left & 81.7270 & 82.0187 & 99.3153 & 89.8421 \\
RESNET50 & Upper Right & 81.7827 & 82.1378 & 99.1784 & 89.8573 \\
VGG16 & Original Images & 97.9944 & 97.9307 & 99.6405 & 98.7782 \\
VGG16 & Bottom Left & 82.5209 & 84.0231 & 96.9531 & 90.0262 \\
VGG16 & Bottom Right & 82.4373 & 84.2223 & 96.4909 & 89.9402 \\
VGG16 & Center & 82.6602 & 84.2804 & 96.7306 & 90.0773 \\
VGG16 & Upper Left & 82.2006 & 83.8275 & 96.7990 & 89.8475 \\
VGG16 & Upper Right & 82.2563 & 83.9578 & 96.6621 & 89.8631 \\
\bottomrule
%\end{tabularx}
\end{tabular}

\end{table}

}

\subsection{CNNs and BreakHis Datasets (Histopathology Imaging)}
\label{BreakHis Results}

\begin{figure}[H]

\hspace{2pt}\includegraphics[width=0.9\linewidth]{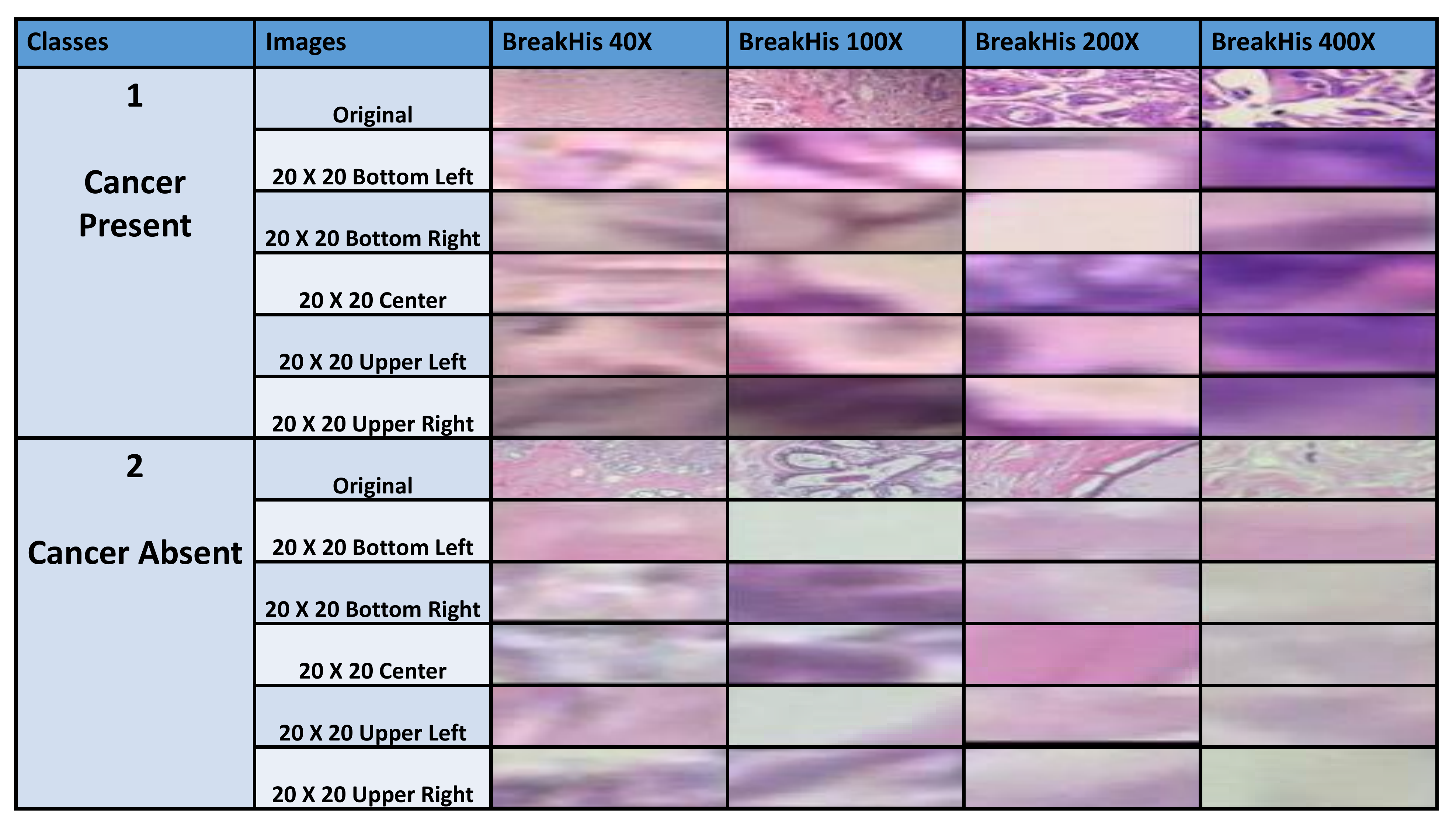} 
\caption {Sample %MDPI: (1) The contents of this figure are not legible. Please replace the image with one of a sufficiently high resolution (min. 1000 pixels width/height, or a resolution of 300 dpi or higher). (2) We moved the figure here after its first citation, please confirm.    Author response: 1. We replaced the image with an image with much better resolution. More importantly, the labels are all in vector graphics so they are clearer to read and will not pixelate at any level of magnification.   2. We confirm. 
 images across different magnification levels of the BreakHis datasets, showing the original images and their 20 $\times$ 20 cropped images. Cropped images show staining protocols that often lack clear histopathological structures.}
\label{Figure BreakHis_Samples}
\end{figure}

In the BreakHis datasets, at a specific magnification level, each image displays tissue structure divided into benign and malignant classes. Images are similar at each magnification level, with consistent color stains and background patterns. Here, lesions may not be centered, and the images contain mainly lesions and background, as shown in Figure~\ref{Figure BreakHis_Samples}.

We evaluated the classification performance of the four CNN models (ResNet50, DenseNet121, InceptionV3, and VGG16) on the BreakHis datasets across the four magnification levels ($40\times$, $100\times$, $200\times$, and $400\times$). The experiment was first performed on the original image dataset and then on the cropped-image datasets. The cropped-image datasets were generated by extracting five 20 $\times$ 20 pixel sections from the original images. Each cropped dataset (bottom-left, bottom-right, center, upper-left, and upper-right) contains images that are too small to contain any meaningful tissue structure for pathology~analysis.

\begin{figure}[H]

\hspace{-20pt}\includegraphics[width=0.9\linewidth]{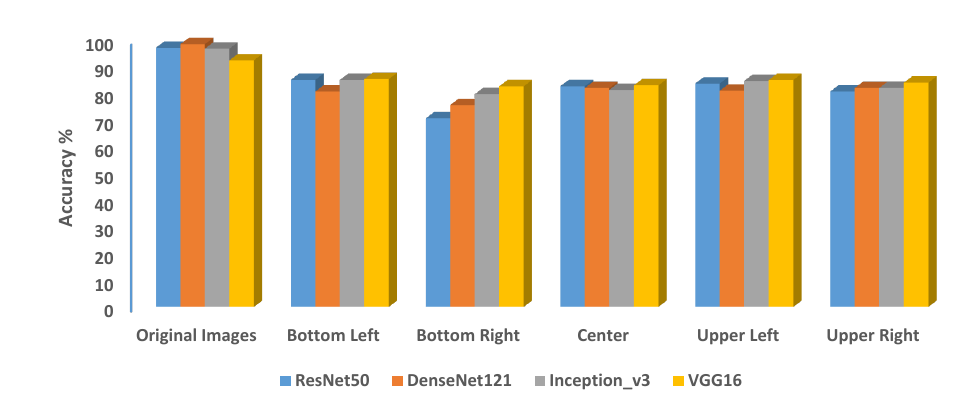}
\caption {The classification accuracy of the BreakHis $40\times$ magnification level of the original images and its corresponding 20 $\times$ 20 cropped sections, showing that CNNs can classify images that contain non-informative medical content.}
\label{Figure BreakHis40}
\end{figure}

\begin{figure}[H]

\hspace{-20pt}\includegraphics[width=0.9\linewidth]{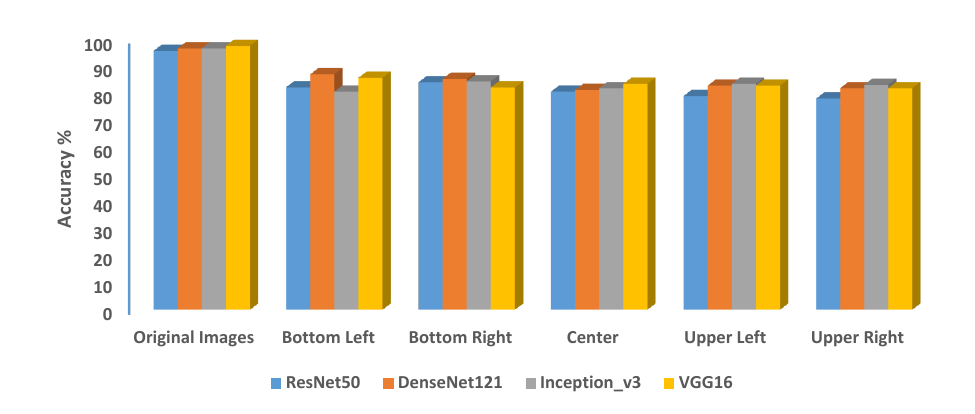}
\caption {All the CNN models can accurately classify the original images as well as all the \mbox{20 $\times$ 20} cropped sections when applied using BreakHis $100\times$, suggesting that neural networks may be learning from staining color or texture pattern not related to the actual disease.}
\label{Figure BreakHis100}
\end{figure}

\begin{figure}[H]

\hspace{-20pt}\includegraphics[width=0.9\linewidth]{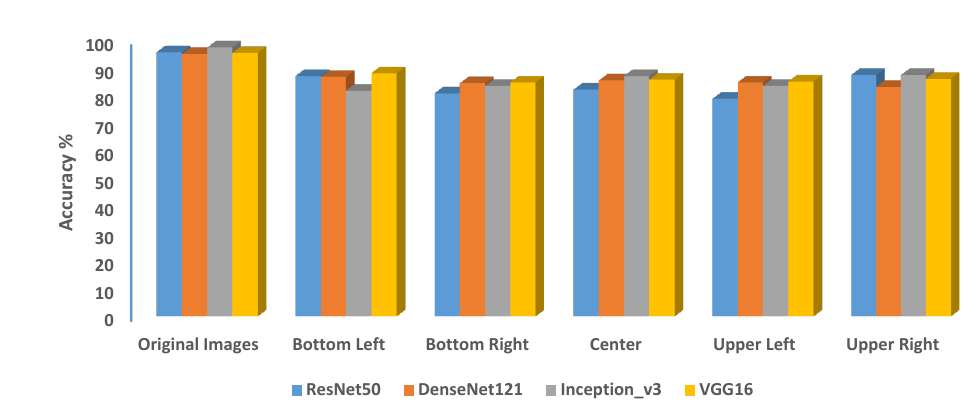}
\caption {The results from BreakHis $200\times$ show that CNN models can classify cancerous or noncancerous tumors from both the original images and their 20 $\times$ 20 cropped sections.}
\label{Figure BreakHis200}
\end{figure}

\begin{figure}[H]

\hspace{-19pt}\includegraphics[width=0.9\linewidth]{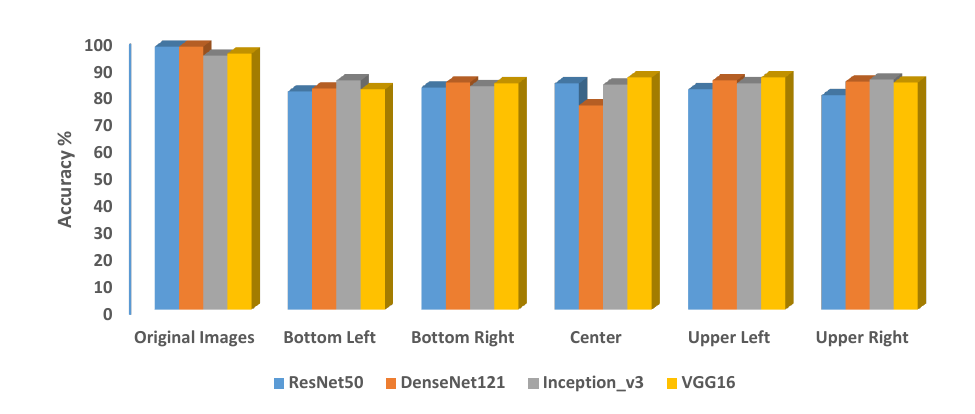}
\caption {The classification accuracy score of all the CNN architectures across BreakHis $400\times$ datasets (the original image and cropped sections) is far higher than accuracy when using only random chance.}
\label{Figure BreakHis400}
\end{figure}

The results revealed the same pattern across all magnification levels. \cref{Figure BreakHis40,Figure BreakHis100,Figure BreakHis200,Figure BreakHis400} show the results from each magnification level of $40\times$, $100\times$, $200\times$, and $400\times$, respectively. Table~\ref{tab:BreakHis_all} summarizes the results for the 40$\times$ magnification.

In these results, the CNN models are shown to be able to classify the original images with an accuracy as high as $95\%$. The results also show that CNNs can identify the classes in the cropped-image datasets with a classification accuracy as high as $88\%$. Achieving a classification accuracy well above that of mere chance, $50\%$, when using the cropped images is unexpected.

The consistent ability of deep neural networks to correctly classify the cropped images across all magnification levels suggests potential bias. Each CNN model can identify class-unique features even when the images contain no visible tissue samples. Deep neural networks may exploit location-based features rather than the generalized learning of disease characteristics. This may occur if certain histological patterns, staining colors, or texture features are consistently present in specific regions of the slides. This raises concerns about the reliability of CNN performance in real-world clinical settings in situations where the tissue presentation of the test distribution may differ from the training distribution.

\vspace{-12pt}

\vspace{-12pt}

\begin{table}[H]
\caption{Summary of the results of the analysis of the BreakHis dataset with 40$\times$ magnification.}
\label{tab:BreakHis_all}
%\renewcommand{\arraystretch}{2.5}
%\begin{tabularx}{\textwidth}{lcCCCC}
\begin{tabular}{lccccc}
\toprule
\textbf{Model %MDPI: We added bold for the header, please confirm.  Author response: We confirm the change.
} & \textbf{Sub Dataset} & \textbf{Accuracy} & \textbf{Recall} & \textbf{F-1} & \textbf{Precision} \\
\midrule
DENSENET121 & Original Images & 97.69 & 97.97 & 98.30 & 98.64 \\
DENSENET121 & Bottom Left & 55.48 & 51.58 & 49.49 & 64.49 \\
DENSENET121 & Bottom Right & 55.35 & 50.48 & 48.87 & 65.09 \\
DENSENET121 & Center & 56.41 & 52.64 & 50.93 & 84.84 \\
DENSENET121 & Upper Left & 55.93 & 52.25 & 50.27 & 85.12 \\
DENSENET121 & Upper Right & 54.74 & 49.95 & 48.53 & 65.76 \\
INCEPTION\_V3 & Original Images & 97.12 & 97.45 & 97.90 & 98.37 \\
INCEPTION\_V3 & Bottom Left & 69.08 & 92.23 & 80.13 & 71.81 \\
INCEPTION\_V3 & Bottom Right & 68.27 & 90.87 & 79.29 & 71.58 \\
INCEPTION\_V3 & Center & 69.14 & 91.79 & 80.05 & 71.89 \\
INCEPTION\_V3 & Upper Left & 68.00 & 90.88 & 79.14 & 71.50 \\
INCEPTION\_V3 & Upper Right & 68.03 & 90.08 & 78.99 & 71.85 \\
RESNET50 & Original Images & 96.06 & 96.63 & 97.12 & 97.66 \\
RESNET50 & Bottom Left & 57.43 & 57.72 & 60.75 & 77.25 \\
RESNET50 & Bottom Right & 56.04 & 54.53 & 58.46 & 77.37 \\
RESNET50 & Center & 57.99 & 58.94 & 61.45 & 77.00 \\
RESNET50 & Upper Left & 57.30 & 57.09 & 60.91 & 77.40 \\
RESNET50 & Upper Right & 55.16 & 52.54 & 57.09 & 77.34 \\
VGG16 & Original Images & 95.15 & 94.80 & 96.35 & 98.01 \\
VGG16 & Bottom Left & 72.18 & 99.79 & 83.22 & 71.38 \\
VGG16 & Bottom Right & 72.66 & 99.67 & 83.45 & 71.79 \\
VGG16 & Center & 72.19 & 99.75 & 83.21 & 71.39 \\
VGG16 & Upper Left & 71.65 & 99.83 & 82.95 & 70.97 \\
VGG16 & Upper Right & 72.22 & 99.71 & 83.22 & 71.43 \\
\bottomrule
%\end{tabularx}
\end{tabular}

\end{table}

\begin{figure}[H]

\includegraphics[width=0.9\linewidth]{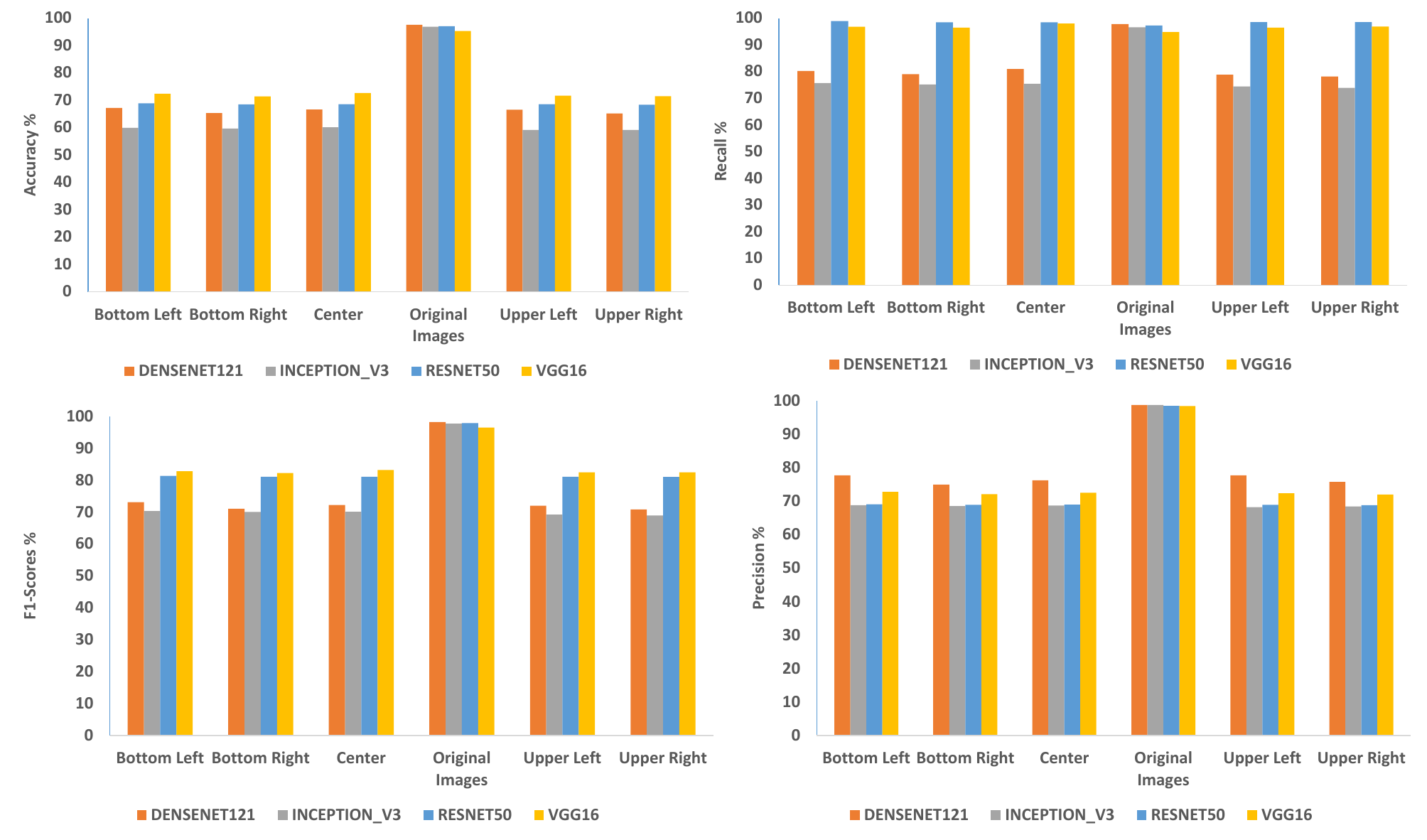}
\caption {The classification accuracy, precision, recall, and F-1 scores for the BreakHis dataset with 40$\times$ magnification when each class contains exactly 405 images.}
\label{BreakHis40_balanced}
\end{figure}

Another experiment was performed to test the effect of class balancing. Figure~\ref{BreakHis40_balanced} shows the classification accuracy, precision, recall and F-1 scores of the 40$\times$ magnification of the BreakHis images. Each class contained 405 images, and the training and testing were carried out using the same settings as the other experiments. As the figure shows, the architectures are still able to identify cancer with an accuracy much higher than mere chance, with the classification accuracy of the original images being close to 100\%, and the background patches provide an accuracy of 70--80\% for most architectures.

\subsection{CNN and ISIC Datasets (Dermoscopic Lesion Skin Imaging)}
\label{ISIC Results}

\begin{figure}[H]
\hspace{2pt}\includegraphics[width=0.9\linewidth]{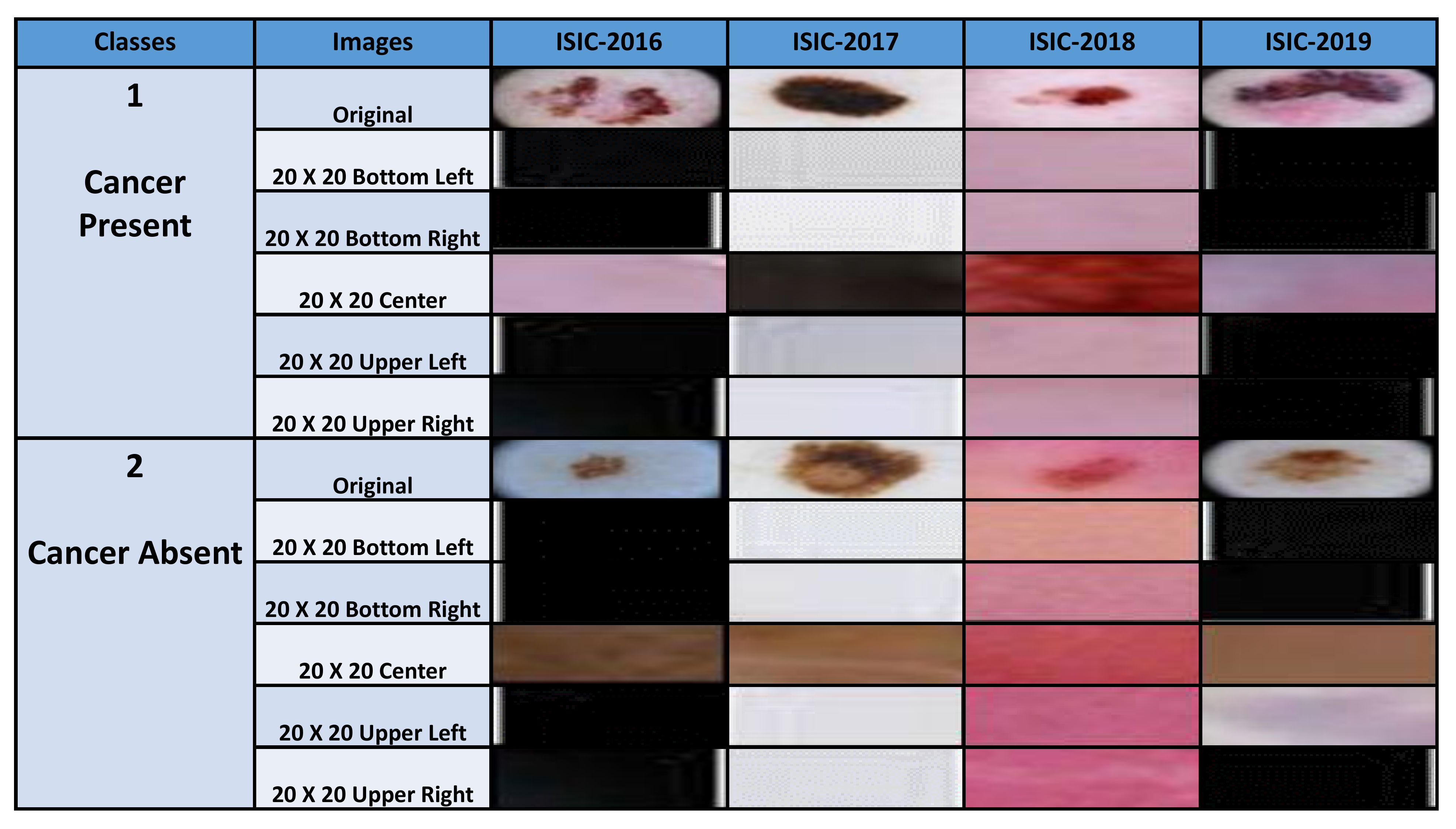}
\caption {Sample %MDPI: Please replace the image with one of a sufficiently high resolution (min. 1000 pixels width/height, or a resolution of 300 dpi or higher).     Author response: Like the other images, we replaced the image with a new figure with a much higher resolution. The labels are with vector graphics so the entire figure is much sharper than before, and will not pixelate at any size of level of magnification.. 
 images across the ISIC datasets, showing the original images and their \mbox{20 $\times$ 20 cropped} images. Mostly, the cropped images contain a black or skin-toned background. Sometimes, center-cropped images include lesion structures that lack meaningful diagnostic features for an expert dermatologist.}
\label{Figure Sample_ISIC}
\end{figure}

The application of deep neural networks for the classification of dermoscopic images is standard practice. However, when we extract 20 $\times$ 20 cropped images from the ISIC original images (four corners and the center), CNNs are not expected to identify cancerous and non-cancerous lesions. These cropped images contain black or skin-toned backgrounds that are not related to the disease pathology. In some cases, the center-cropped images contain pigmentation that is diagnostically meaningless to an expert dermatologist. Figure~\ref{Figure Sample_ISIC} shows sample images of the original images and their corresponding 20 $\times$ 20 cropped images across the ISIC datasets. 

Despite this, the results of the experiments across ISIC datasets (ISIC-2016 to ISIC-2019) show that CNNs can distinguish malignant from benign cases in these cropped images. In some cases, CNNs achieved a marginally higher accuracy score on the cropped images than on the original images. 

\begin{figure}[H]
\hspace{-21pt}\includegraphics[width=1.0\linewidth]{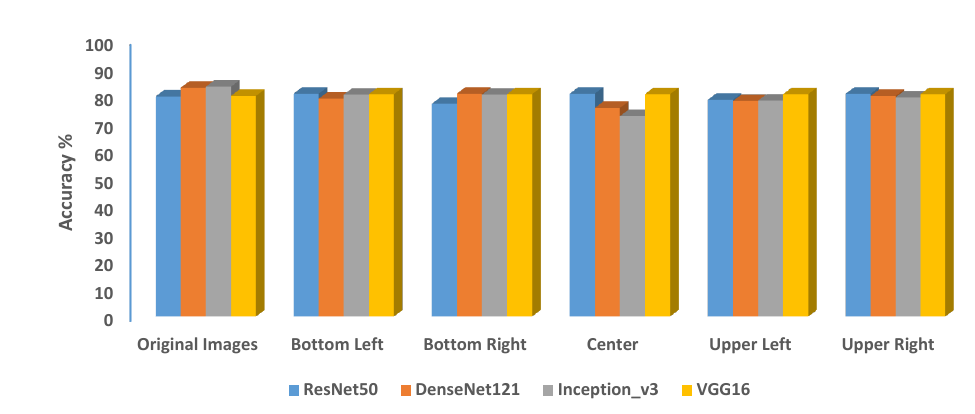}
\caption {All CNN models can classify the ISIC\_2017 cropped-image datasets with an accuracy higher than the 50\% of mere chance. VGG16 consistently performs slightly better on the cropped images than on the original images.}
\label{Figure ISIC-2017}
\end{figure}

For example, VGG16 exhibited a consistent accuracy performance of $\sim$80.5\% across all cropped images datasets of ISIC-2017, as shown in Figure%MDPI: There is mentioned Figure 18 before Figure 17, figures should be cited in numerical order, please check and revise.    Author response: We changed the order of the figures by moving them inside the latex file.
~\ref{Figure ISIC-2017}. This performance is slightly higher than the performance of $\sim$79.83\% it achieved on the original images. ResNet50 also showed a similar trend by achieving a consistent accuracy score of $\sim$80.67\% for bottom-left, center-, and upper-right cropped images, which is slightly higher than the accuracy it achieved on the original images.

\begin{figure}[H]

\hspace{-20pt}\includegraphics[width=1.0\linewidth]{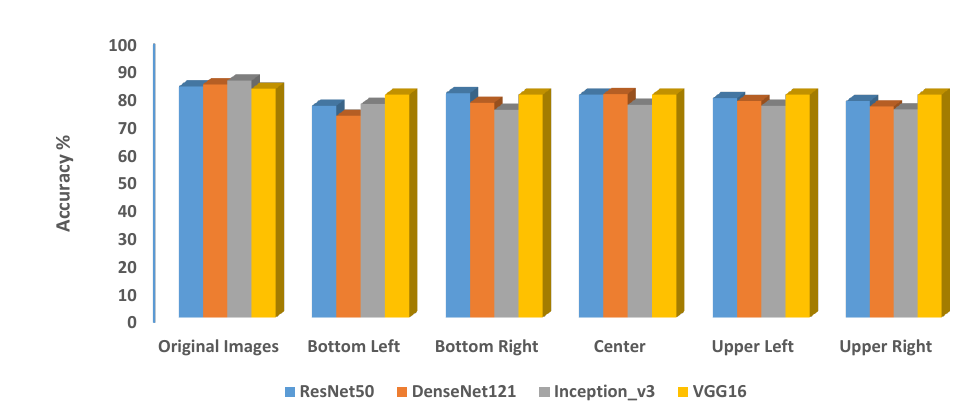}
\caption {The %MDPI: We moved the Figures 17 and 18 here after its first citation, please confirm.   Author response: We confirm.
 classification accuracy of the ISIC\_2016 original image dataset alongside its \mbox{20 $\times$ 20 cropped} sections; VGG16 exhibited consistent accuracy performance across all cropped datasets, suggesting that neural networks may be learning from background artifacts during training.}
\label{Figure ISIC-2016}
\end{figure}

As shown in Figure~\ref{Figure ISIC-2016}, VGG16 also achieved a consistent accuracy performance of $\sim$80.21\% when applied to all the cropped-image datasets of ISIC-2016, in contrast to its performance on the ISIC-2017 dataset, where it achieved a slightly lower performance than that achieved on the original images. 

\begin{figure}[H]
\hspace{-20pt}\includegraphics[width=1.0\linewidth]{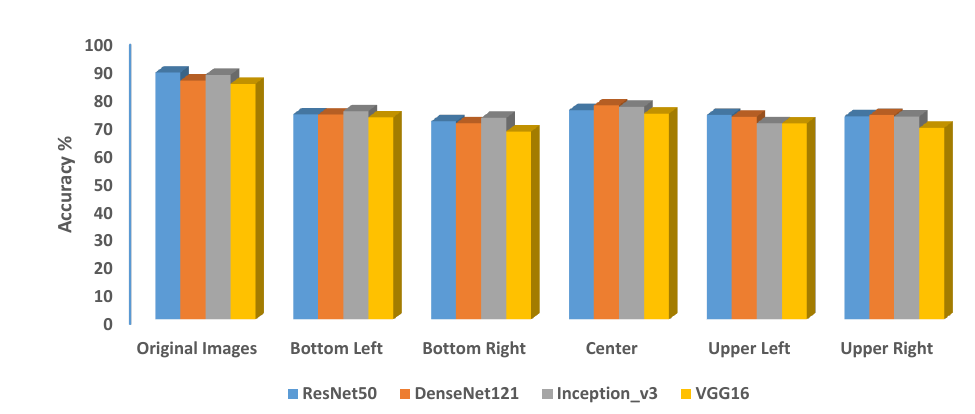}
\caption {The %MDPI: Please cite the figure in the text and ensure the first citation of each figure appears in numerical order.   Author response: The figure is cited in Section 3.3.
 classification accuracy scores on the ISIC\_2018 original images and the cropped sections show that CNN models can also identify skin cancer from a small fraction of the full dermatoscope image containing only skin-toned background.}
\label{Figure ISIC-2018}
\end{figure}

Another dataset we tested is the ISIC-2018. The results also reveal the ability of CNN models to identify classes from the cropped images. However, the margin between the performance across models on the original images and the cropped-image datasets increased. This suggests little resistance to potential bias compared to the results from previous years (ISIC-2016 and ISIC-2017). For example, Figure~\ref{Figure ISIC-2018} shows that deep neural networks can classify both the original and cropped images with accuracy as high as $\sim$88.29\% and $\sim$76.57\%, respectively.

\vspace{-12pt}

\begin{figure}[H]
\hspace{-20pt}\includegraphics[width=1.0\linewidth]{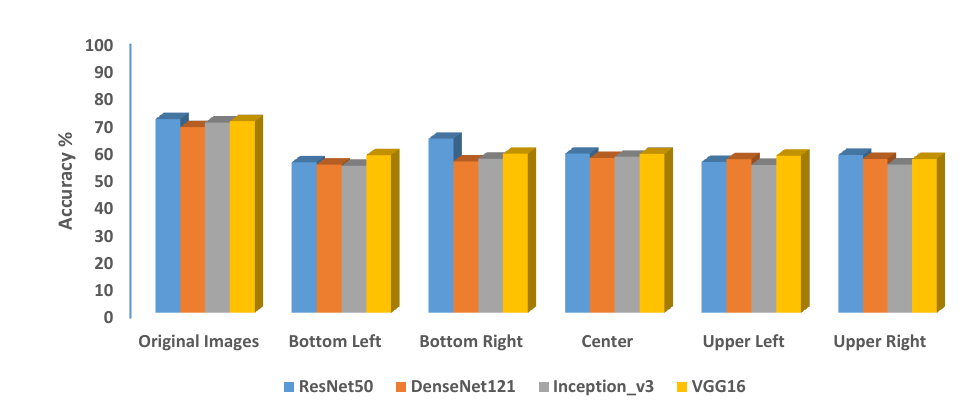}
\caption {Classification accuracy scores on the ISIC\_2019 original images and cropped sections. Performance reveals a significant drop compared to previous years, but CNN can still classify cropped images with an accuracy slightly higher than mere chance of 50\%.}
\label{Figure ISIC-2019}
\end{figure}

{
Table~\ref{isic_2018_balanced} shows the classification accuracy. precision, recall and F-1 scores of the ISIC-2018 dataset when the classes are balanced. As before, different parts of each image are used. As the table shows, in all cases the classification accuracy is higher than mere chance-based accuracy, even when using sub-images with no medical information. The classification accuracy is consistent across different parts of the images at around 63\%.

When the experiment was performed on the ISIC-2019 dataset, the results reveal a significant drop in overall performance. All CNN architectures achieved lower accuracy scores compared to previous years, both on the original and cropped-image datasets. The performance of the models on the original images is between $\sim$70.93\% and $\sim$67.93\%, with these results achieved by ResNet50 and DenseNet121, respectively. Most CNN models showed a moderate performance when applied to the cropped-image datasets. In most cases, performance was only slightly higher than that of mere chance, 50\%. InceptionV3 achieved the lowest performance of $\sim$54.33\% and ResNet50 achieved the maximum performance of $\sim$63.75\%, as shown in {Figure}~\ref{Figure ISIC-2019}. This shows that these results might be less subjected to bias.

\begin{table}[H]
\caption{Accuracy, precision, recall, and F-1 results of the analysis of the balanced ISIC-2018 dataset.}
\label{isic_2018_balanced}
\small
%\renewcommand{\arraystretch}{2.5}
%\begin{tabularx}{\textwidth}{lcCCCC}
\begin{tabular}{lccccc}
\toprule
\textbf{Model %MDPI: We added the bold for the header, please confirm.   Author response: We confirm.
} & \textbf{Sub Dataset }& \textbf{Accuracy} & \textbf{Recall} & \textbf{F-1 }& \textbf{Precision} \\
\midrule
DENSENET121 & Original Images  & 85.8372 & 85.5634 & 81.9552 & 78.7535 \\
DENSENET121 & Bottom Right & 62.8590 & 4.4366 & 7.4707 & 28.3412 \\
DENSENET121 & Bottom Left & 62.7399 & 3.9437 & 6.8230 & 35.0489 \\
DENSENET121 & Center & 61.2442 & 6.5141 & 10.0572 & 64.2105 \\
DENSENET121 & Upper Left & 62.9782 & 4.6479 & 7.9069 & 57.9132 \\
DENSENET121 & Upper Right & 62.8855 & 4.7887 & 7.9559 & 42.9699 \\
INCEPTION\_V3 & Original Images  & 84.8312 & 84.7887 & 80.8230 & 77.6369 \\
INCEPTION\_V3 & Bottom Right & 62.6208 & 2.8169 & 4.5410 & 41.0638 \\
INCEPTION\_V3 & Bottom Left & 62.7796 & 2.6761 & 4.5365 & 24.9397 \\
INCEPTION\_V3 & Center & 62.2766 & 2.3592 & 4.1722 & 19.4480 \\
INCEPTION\_V3 & Upper Left & 62.7929 & 2.5000 & 4.2595 & 29.0031 \\
INCEPTION\_V3 & Upper Right & 62.7929 & 2.6761 & 4.5215 & 25.6140 \\
RESNET50 & Original Images  & 85.1621 & 83.2042 & 80.8063 & 78.8106 \\
RESNET50 & Bottom Right & 63.0443 & 6.9014 & 11.4470 & 58.2067 \\
RESNET50 & Bottom Left & 63.0841 & 5.8451 & 9.9479 & 44.7500 \\
RESNET50 & Center & 61.7737 & 1.6197 & 2.8298 & 13.2013 \\
RESNET50 & Upper Left & 63.6929 & 8.2042 & 13.4575 & 57.7125 \\
RESNET50 & Upper Right & 63.6664 & 7.9930 & 13.1650 & 68.3775 \\
VGG16 & Original Images  & 82.9385 & 82.4296 & 78.4074 & 74.9589 \\
VGG16 & Bottom Right & 63.0311 & 7.9930 & 11.7898 & 42.2358 \\
VGG16 & Bottom Left & 63.0179 & 7.2183 & 10.9649 & 32.5476 \\
VGG16 & Center & 62.8326 & 6.6549 & 11.3717 & 62.9896 \\
VGG16 & Upper Left & 63.3091 & 7.8521 & 11.8730 & 63.5431 \\
VGG16 & Upper Right & 63.2694 & 7.6056 & 11.5838 & 53.5498 \\
\bottomrule
%\end{tabularx}
\end{tabular}

\end{table}

}

{
Table~\ref{isic_2019_balanced} shows the accuracy, precision, recall, and F-1 results when the classes are balanced. As with the previous experiments with that dataset, the classification accuracy using the background sub-images with no medical information, in most cases, was approximately the same as the expected accuracy when using only chance. These results are not necessarily surprising, given the results shown in Figure~\ref{Figure ISIC-2019}.

\begin{table}[H]
\caption{Accuracy, precision, recall, and F-1 results of the analysis of the balanced ISIC-2019 dataset.}
\label{isic_2019_balanced}
\footnotesize
%\renewcommand{\arraystretch}{2.5}
%\begin{tabularx}{\textwidth}{lcCCCC}
\begin{tabular}{lccccc}
\toprule
\textbf{Model %MDPI: We added the bold for the header, please confirm.   Author response: We confirm the change.
} & \textbf{Sub Dataset }& \textbf{Accuracy} & \textbf{Recall} &\textbf{ F-1} & \textbf{Precision} \\
\midrule
DENSENET121 & Original Images & 0.68 & 0.53 & 0.78 & 0.63 \\
DENSENET121 & Bottom Left & 0.48 & 0.38 & 0.80 & 0.52 \\
DENSENET121 & Bottom Right & 0.48 & 0.38 & 0.82 & 0.52 \\
DENSENET121 & Center & 0.48 & 0.30 & 0.40 & 0.34 \\
DENSENET121 & Upper Left & 0.48 & 0.38 & 0.82 & 0.52 \\
DENSENET121 & Upper Right & 0.49 & 0.39 & 0.84 & 0.53 \\
INCEPTION\_V3 & Original Images & 0.68 & 0.53 & 0.80 & 0.64 \\
INCEPTION\_V3 & Bottom Left & 0.41 & 0.35 & 0.84 & 0.50 \\
INCEPTION\_V3 & Bottom Right & 0.43 & 0.36 & 0.85 & 0.51 \\
INCEPTION\_V3 & Center & 0.46 & 0.37 & 0.78 & 0.50 \\
INCEPTION\_V3 & Upper Left & 0.45 & 0.37 & 0.85 & 0.52 \\
INCEPTION\_V3 & Upper Right & 0.44 & 0.37 & 0.85 & 0.51 \\
RESNET50 & Original Images & 0.67 & 0.51 & 0.84 & 0.63 \\
RESNET50 & Bottom Left & 0.55 & 0.40 & 0.67 & 0.50 \\
RESNET50 & Bottom Right & 0.54 & 0.40 & 0.66 & 0.50 \\
RESNET50 & Center & 0.52 & 0.31 & 0.31 & 0.31 \\
RESNET50 & Upper Left & 0.55 & 0.40 & 0.67 & 0.51 \\
RESNET50 & Upper Right & 0.55 & 0.40 & 0.66 & 0.50 \\
VGG16 & Original Images & 0.61 & 0.46 & 0.90 & 0.61 \\
VGG16 & Bottom Left & 0.57 & 0.41 & 0.58 & 0.48 \\
VGG16 & Bottom Right & 0.56 & 0.41 & 0.62 & 0.49 \\
VGG16 & Center & 0.64 & 0.37 & 0.05 & 0.09 \\
VGG16 & Upper Left & 0.56 & 0.41 & 0.64 & 0.50 \\
VGG16 & Upper Right & 0.55 & 0.41 & 0.67 & 0.51 \\
\bottomrule
%\end{tabularx}
\end{tabular}

\end{table}

}

\vspace{-15pt}

The results across all datasets (ISIC-201 to ISIC-2017) indicate that neural networks may be learning from irrelevant background features in the original images. This may influence their ability to classify images lacking disease-related features. In later years (ISIC-2018 and ISIC-2019), the noticeable drop in the performance of CNN on the cropped-image datasets points to reduced susceptibility to bias compared to the results from previous years. This indicates that the bias in the performance of deep neural networks may be related to the complexity of biomedical image datasets.

\begin{figure}[H]

\includegraphics[width=0.8\linewidth]{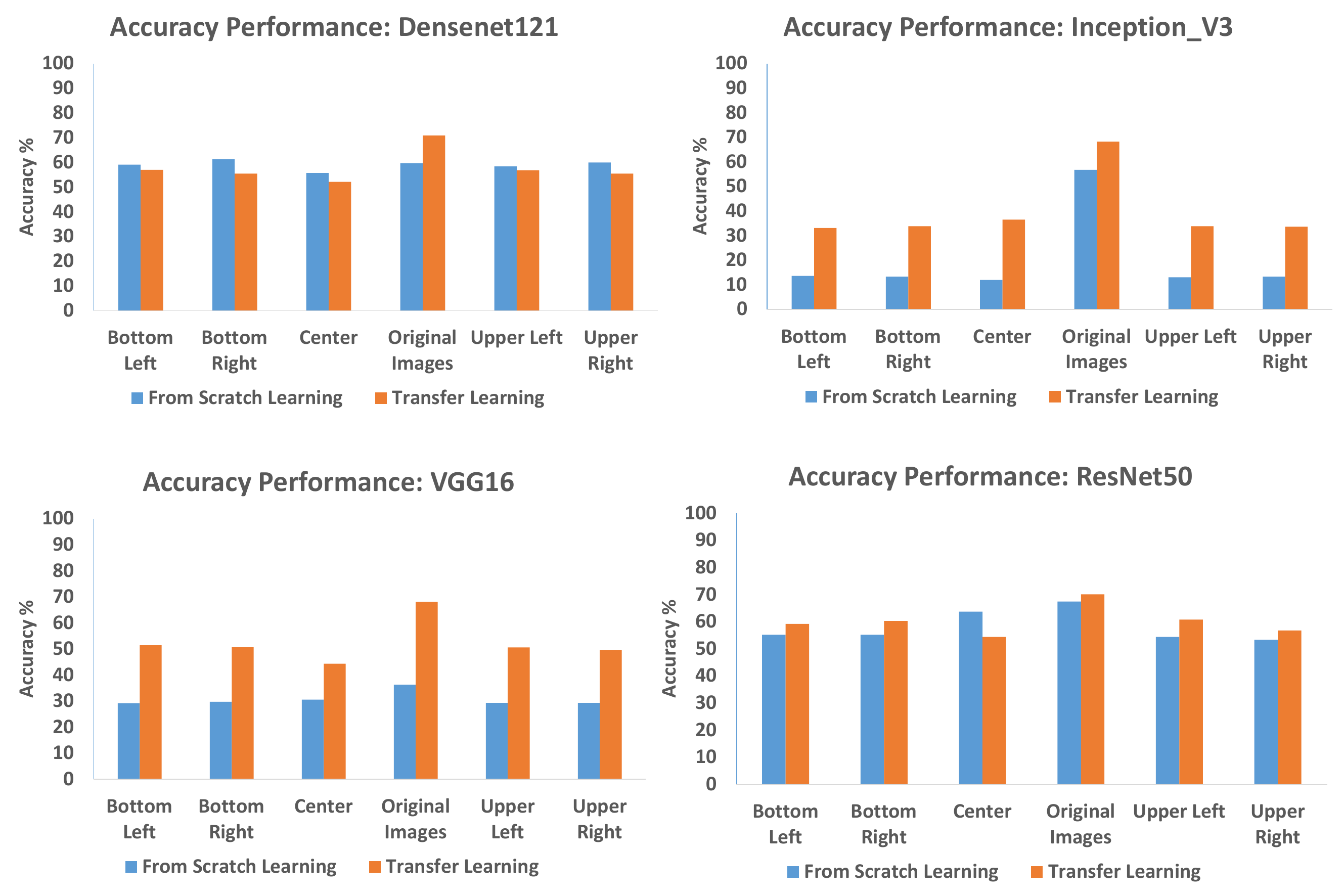}
\caption {Classification accuracy of ISIC-2016 using different parts of the image with or without using transfer learning. The number of epochs was set to 15 in both cases.}
\label{isic_trabsfer_learning}
\end{figure}

We also used ISIC to test the effect of transfer learning. Transfer learning is the standard practice for training deep neural networks, and therefore the experiments above were performed with transfer learning. However, to test if the bias derived from the practice of transfer learning, we also tested ISIC-2016 with and without transfer learning. That was achieved by using 15 epochs when training with or without transfer learning to ensure that the only difference between the experiments was the use of transfer learning. Figure~\ref{isic_trabsfer_learning} shows the classification accuracy of the different parts with or without transfer learning. The analysis was carried out for the different CNNs architectures.

As the figure shows, transfer learning performed better when classifying the original images. That is expected, as transfer learning is known to improve the training; therefore, the common practice is to use transfer learning.

\begin{figure}[H]

\includegraphics[width=1.0\linewidth]{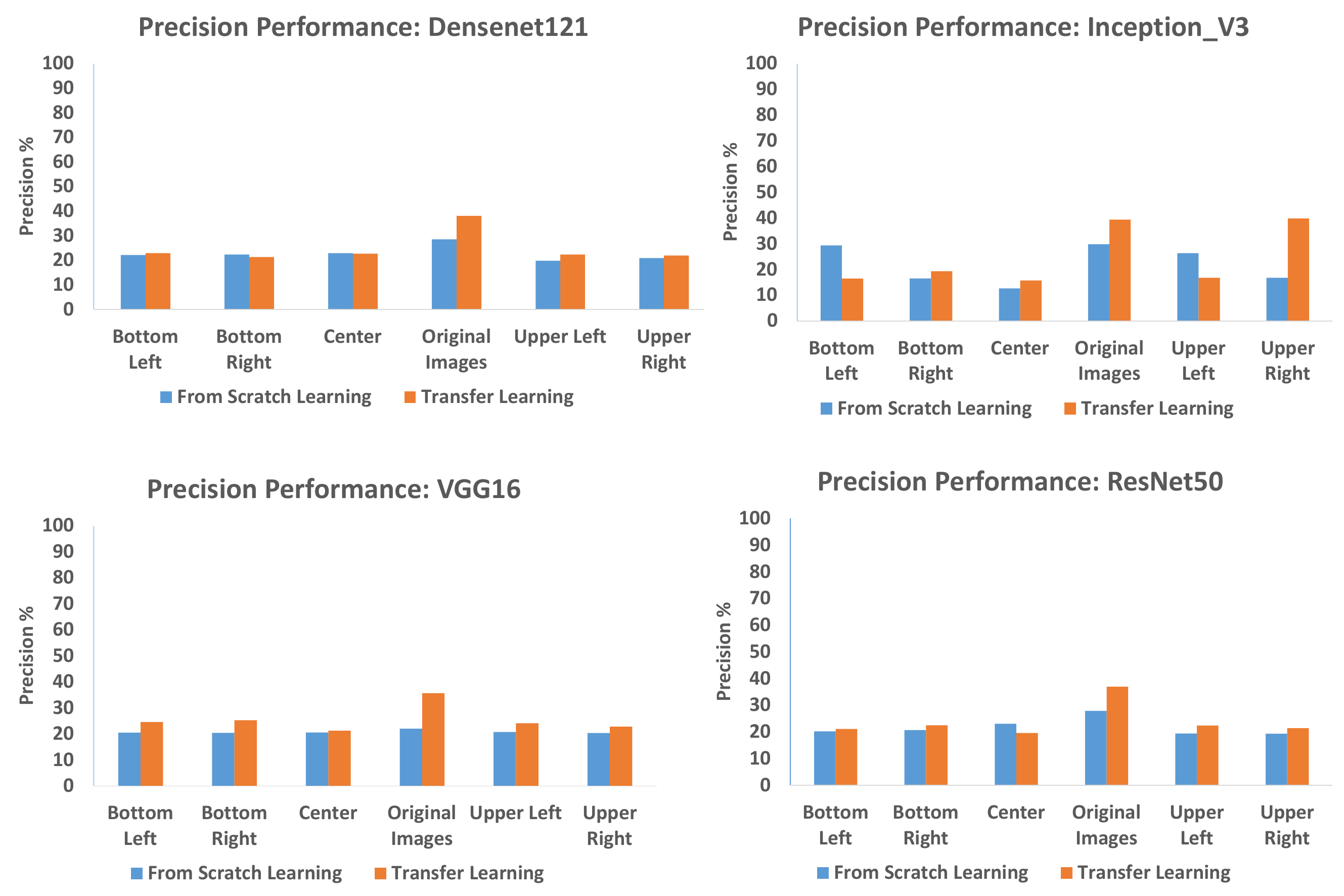}
\caption {Precision observed when using different deep neural network architectures when using ISIC-2016 with different parts of the image, with and without using transfer learning.}
\label{isic_trabsfer_learning_precision}
\end{figure}

\begin{figure}[H]

\includegraphics[width=1.0\linewidth]{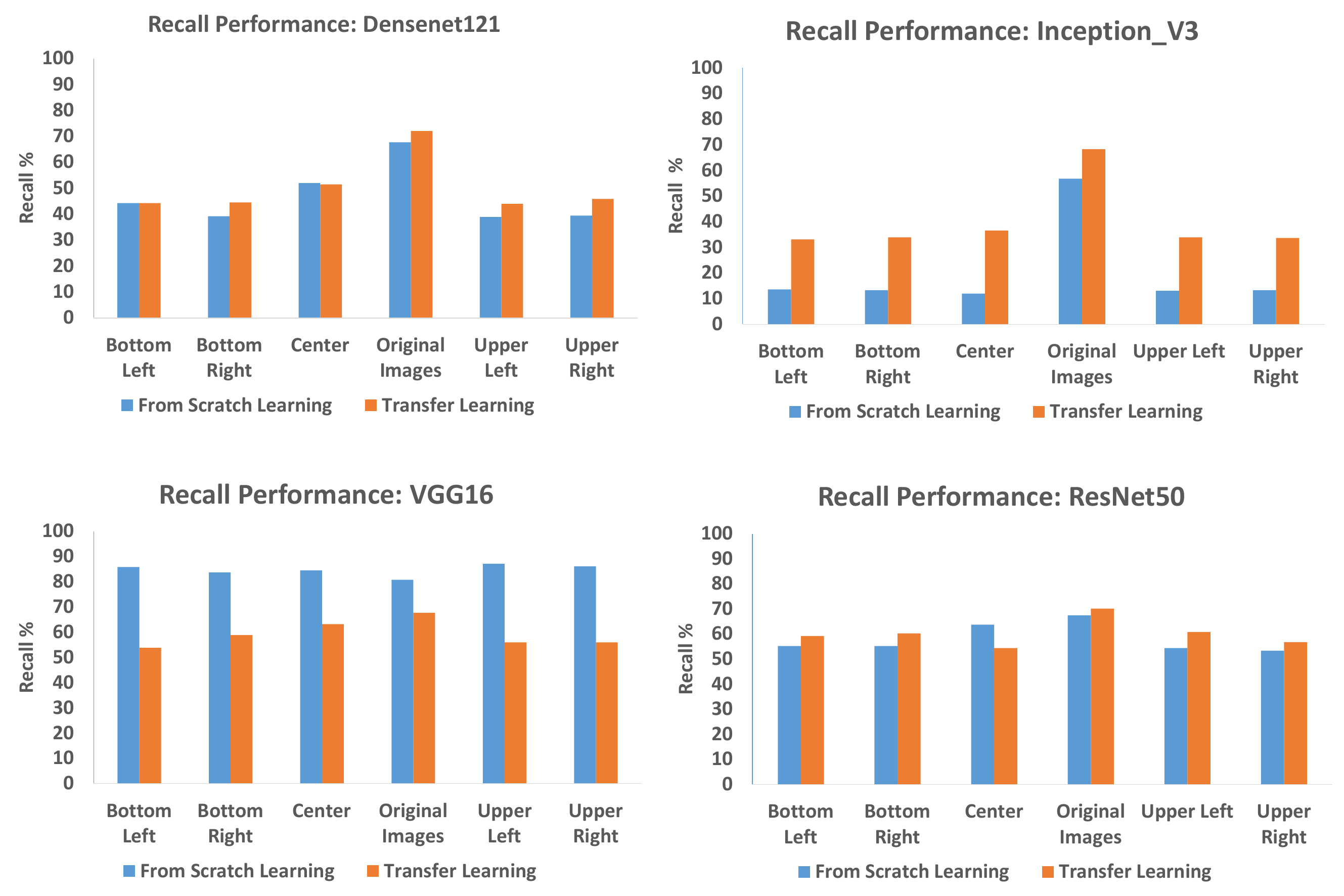}
\caption {Recall of different architectures when using ISIC-2016 with different parts of the image, with or without using transfer learning.}
\label{isic_trabsfer_learning_recall}
\end{figure}

\begin{figure}[H]

\includegraphics[width=1.0\linewidth]{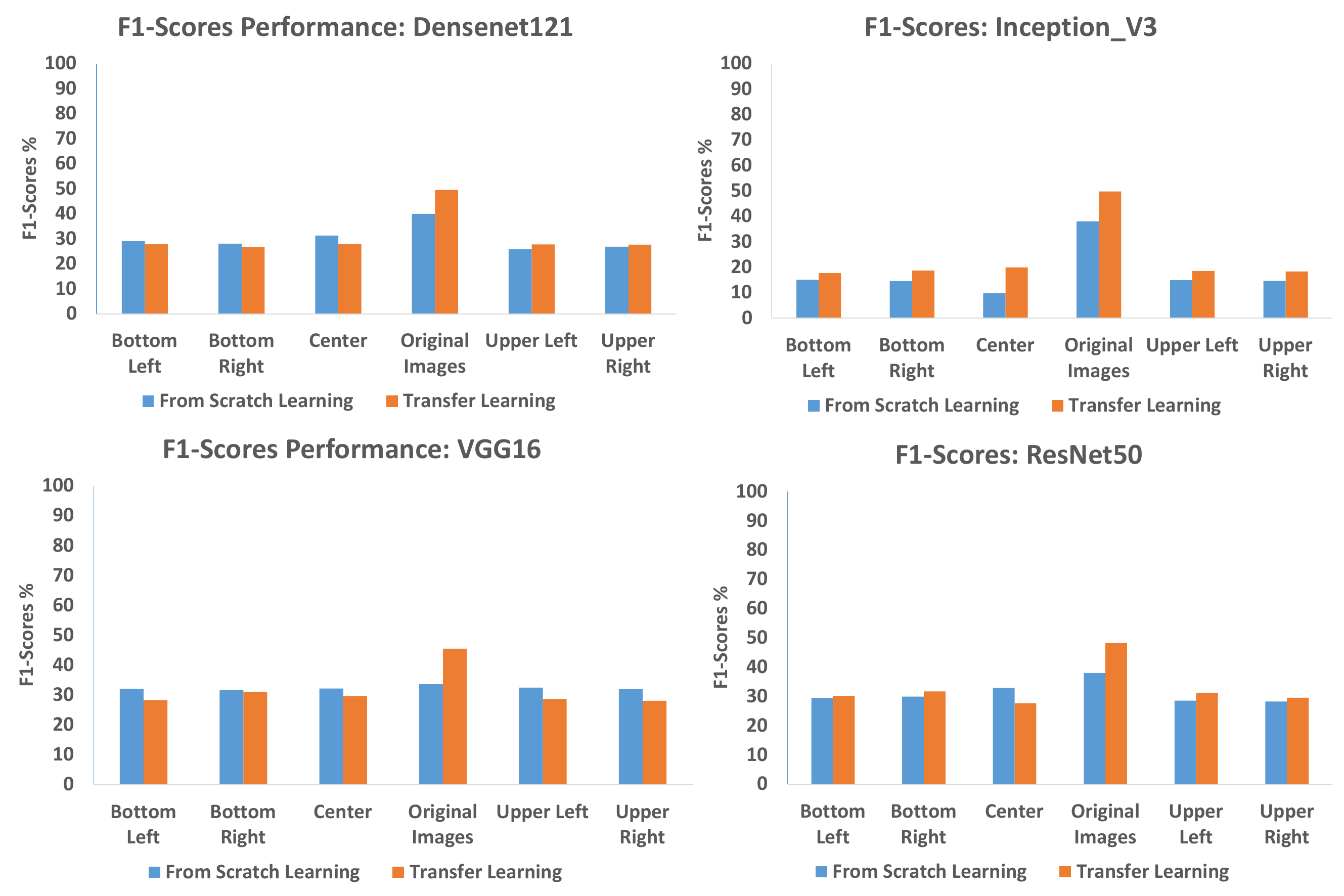}
\caption {F-1 of different architectures when using ISIC-2016 with different parts of the image, with or without using transfer learning.}
\label{isic_trabsfer_learning_f1}
\end{figure}

Interestingly, with the exception of VGG-16, the bias did not increase substantially when using transfer learning, and in the case of Inception V3 it even decreased. That shows that the bias is not driven by the information that is present in common natural images, but is specific to the image dataset %MDPI: We merged the citation, please confirm.   Author response: We confirm.
. \cref{isic_trabsfer_learning_precision,isic_trabsfer_learning_recall,isic_trabsfer_learning_f1} show the precision, recall, and F-1 scores when using different parts of the images and different architectures. 

\begin{figure}[H]

\includegraphics[width=0.9\linewidth]{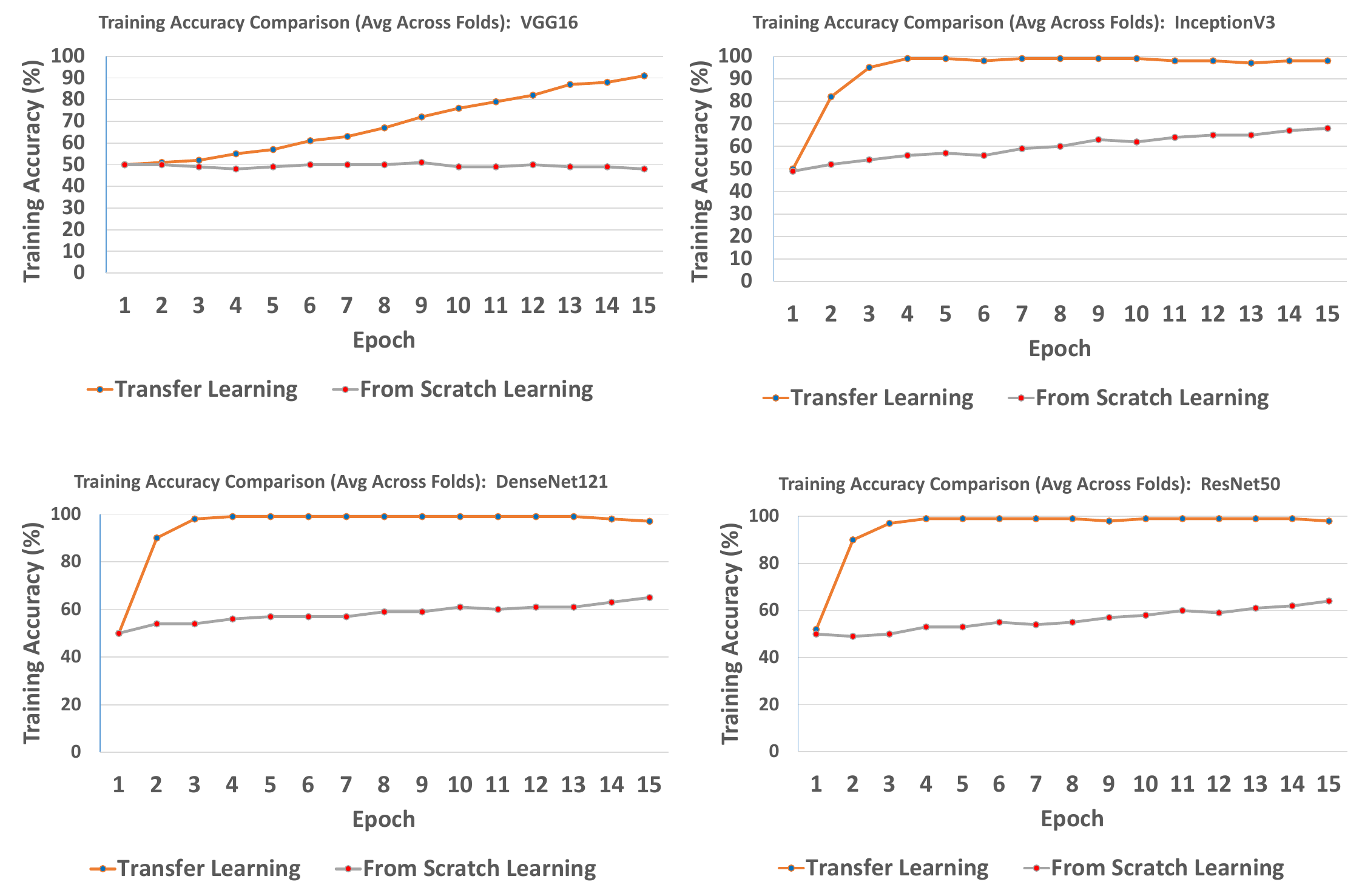}
\caption {Learning as a function of the number of epochs for different CNN architectures.}
\label{learning_curve}
\end{figure}

Figure~\ref{learning_curve} shows the learning with different numbers of epochs for the different architectures. As expected, transfer learning makes the learning substantially faster compared to starting with random weights. For the deeper architectures such as Densenet-121 and Resnet-50, the difference between transfer learning and random weights is even more substantial. 

\subsection{CNN and Breast Histopathology Images for Invasive Ductal Carcinoma (IDC) Dataset (Histopathology Imaging)}
\label{IDC Results}

\textls[-10]{The IDC datasets contain H\&E-stained breast tissue slides of histopathology image tiles. Each image is labeled either cancerous or noncancerous, making them suitable for binary~classification. }

To test for bias, we further cropped each original image into 20 $\times$ 20 patches from five regions. These cropped-image datasets contain only plain tissues or small fractions of breast tissues, instead of the full cancer structure.  The information contained in these cropped images is not sufficient for accurate diagnostics. % We do not expect deep neural networks to predict the class, since they lack the lesion structure needed for cancer detection. 

\begin{figure}[H]

\hspace{-21pt}\includegraphics[width=1.0\linewidth]{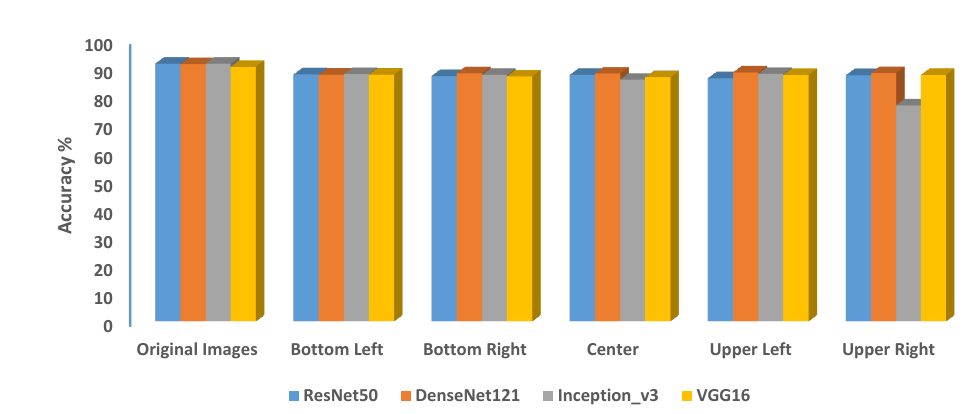}
\caption {The classification accuracy scores on the breast histopathology original images and cropped images show that the models can accurately detect the classes, whether IDC-positive or IDC-negative, on cropped images containing little or no medical content.}
\label{Figure IDC}
\end{figure}

An accuracy score of above 89\% was observed with all deep neural networks on the Breast Histopathology Image (IDC) original image dataset. However, all models also achieved far higher than the random accuracy across all cropped 20 $\times$ 20 subset datasets, with ResNet50 having showing the lowest performance of $\sim$85\%, as shown in Figure~\ref{Figure IDC}. The ability of these CNN models to successfully identify the classes in the cropped images that do not contain sufficient clinical information shows that the results observed with the full images could be biased. These results are aligned with other histopathology image datasets, such as PathMNIST and BreakHis, as shown above.

\vspace{-3pt}

\section{Conclusions}
\label{conclusion}

CNNs have become a widely adopted tool for automatic pathology. Due to the availability of easy-to-use libraries, CNNs have become powerful tools for cancer pathology. This study highlights a weakness in the common practice of the use of CNNs for biomedical image analysis, unmasking their vulnerability to bias when applied to cancer diagnostics.

We examined whether deep neural networks learn solely from meaningful diagnostic features, or also rely on background details/dataset-specific artifacts. To investigate this, we evaluated the performance of four commonly used CNN models across thirteen diverse biomedical datasets. These datasets are widely used by machine learning experts and the medical AI community, representing different types of cancer and imaging modalities. We then compared the performance of each CNN model on the original image to that observed when using 20 $\times$ 20 pixel regions. These cropped images contain little or no medical content useful for meaningful diagnosis. 

Surprisingly, the results across all datasets showed that CNNs consistently achieved an accuracy higher than random chance on the cropped image. In some cases, CNNs achieved a matching or better performance than that achieved the original images, even when the images lacked relevant diagnostic features. { In other cases, the original images allowed for better classification accuracy compared to the small background sub-images, which is expected given that the original images are larger and therefore contain more information that allows for classification, including the medically relevant parts of the images. However, the fact that the background sub-images provided a classification accuracy that is better than random chance, despite containing no medically relevant information, shows that the CNN models did not solely learn from cancer-specific visual features during training. This effect cuts across all types of cancer and imaging modalities, underscoring that the bias issue is both widespread and modality-independent.}

The findings show that CNNs make use of information that is not necessarily clinical to identify between the classes. Therefore, the results observed by applying CNNs to cancer pathology might be substantially biased, and do not necessarily represent the real ability of the CNN to identify cancer. While the use of CNN models on benchmark datasets appears promising, such performance may not reflect true pathology interpretation. This may pose a risk in critical medical contexts, such as cancer diagnosis, where incorrect diagnoses can have severe consequences. % By unmasking this weakness, this study contributes to the development of bias-aware medical diagnostic architectures. This will help ensure that deep learning tools for medical diagnosis learn only from disease-specific features. In the future, this will lead to building safer, more interpretable, and more generalizable deep neural networks for cancer diagnosis. 

%\section{Limitations And Future Work}
%\label{limitations}

Although this study exposes potential bias in the common practice of using deep neural networks for biomedical image analysis, it does not scientifically investigate the root causes of the bias. The analysis is limited to four CNN architectures and thirteen widely used biomedical image datasets. Although these architectures and benchmarks are commonly used, it is definitely possible that some benchmarks and architectures are not biased. The size of the sub-images is 20 $\times$ 20, which is based on previous experiments on datasets from other domains~\citep{dhar2021evaluation,dhar2022systematic,shamir2008evaluation}. While other sizes could provide different results, the fact that classification accuracy can be observed with no medical information indicates the possible presence of bias. Since all images are processed in the same way, artifacts that are derived from the image processing should affect all classes equally, and therefore should not lead to differences between the classes.

Since CNNs are not intuitive, and work largely as a ``black box'', interpretation of the sources of bias is not trivial, and requires understanding the conditions under which the benchmark datasets were prepared. That is, while the outcome of this experiment across all datasets revealed bias, the study did not explore why such bias exists or how it can be~addressed.

%This comparative evaluation reveals the presence of bias, which is based on how well CNN models can classify cropped images with minimal or no medical content. 

Future research work will focus on profiling not merely the presence of bias, but also the reasons for it. That will involve collecting biomedical image datasets instead of relying on publicly available datasets.  Control over the image acquisition process will allow for the proper documentation of relevant clinical variables such as temperature, camera settings, staining protocols, lighting, and acquisition devices during image processing and clinical workflow. This creates the possibility for researchers to systematically analyze how these clinical and acquisition factors contribute to CNN model bias. 

Furthermore, this will serve as a foundation for developing methods for reducing or totally mitigating such bias in the use of deep neural networks for cancer diagnosis. That will be achieved using explainable AI tools and engineered features such as textures or polynomial decomposition of the images, as well as common tools such as saliency maps.

%\section*{Ethics statement}
%The research was done using publicly available data. By design, no new data were collected in this study, and therefore no human subjects were involved.

%\section*{Declaration of competing interest}
%The authors declare no known conflict of interest or competing financial interests. .

\vspace{6pt}
\authorcontributions{~} Conceptualization: M.O., E.M., A.D., and L.S.; validation: M.O.;  methodology: M.O.; software: M.O. and E.M., investigation: M.O., E.M., A.D., and L.S.; resources, L.S.; writing: M.O., E.M., A.D., and L.S.; visualization, M.O.; supervision, L.S. All authors have read and agreed to the published version of the manuscript.

%MDPI: For research articles with several authors, the following statements should be used “Conceptualization, X.X. and Y.Y.; methodology, X.X.; software, X.X.; validation, X.X., Y.Y. and Z.Z.; formal analysis, X.X.; investigation, X.X.; resources, X.X.; data curation, X.X.; writing—original draft preparation, X.X.; writing—review and editing, X.X.; visualization, X.X.; supervision, X.X.; project administration, X.X.; funding acquisition, Y.Y. All authors have read and agreed to the published version of the manuscript.” Author response: A statement has been added.

\funding{ %MDPI: Please add: This research received no external funding or This research was funded by [name of funder] grant number [xxx] And The APC was funded by [XXX]. Information regarding the funder and the funding number should be provided. Please check the accuracy of funding data and any other information carefully.  Author response: A funding statement has been added.
The research was funded in part by the USA National Science Foundation grant 2148878 and by the Johnson Cancer Research Center (JCRC).
}

\dataavailability{The paper is based on previously published data, and therefore all datasets used in this paper are publicly available in the cited papers. MedMNIST+, PathMNIST, DermaMNIST, BreastMNIST, and NoudleMNIST can be publicly accessed at %MDPI: Please add the access date (format: Date Month Year), e.g., accessed on 1 January 2020. the same below.
 \url{https://zenodo.org/records/10519652}. BreakHis is publicly available at \url{https://www.kaggle.com/datasets/ambarish/breakhis}. ISIC-2016, ISIC-2017, ISIC-2018, and ISIC-2019 are publicly available at \url{https://challenge.isic-archive.com/}. IDC can be accessed at \url{https://www.kaggle.com/datasets/paultimothymooney/breast-histopathology-images}.}

\acknowledgments{We would like to thank the three anonymous reviewers for the insightful comments. %  \hl{The research was funded in part by National Science Foundation grant 2148878} %MDPI: Please check if this should be moved to Funding part.  Author response: We removed it as suggested.
. }
\conflictsofinterest{The %MDPI: We moved from acknowledgments to here, please confirm.  Author response: We confirm.
 authors declare no conflicts of interest.
}

\begin{adjustwidth}{-\extralength}{0cm}
\reftitle{References}

\end{adjustwidth}

\begin{adjustwidth}{-\extralength}{0cm}
%\centering %% If there is a figure in wide page, please release command \centering
\PublishersNote{}
\end{adjustwidth}


\begin{thebibliography}{999}

\bibitem[Echle et~al.(2021)Echle, Rindtorff, Brinker, Luedde, Pearson, and
  Kather]{echle2021deep}
Echle, A.; Rindtorff, N.T.; Brinker, T.J.; Luedde, T.; Pearson, A.T.; Kather,
  J.N. %MDPI: Please do not change the reference format. Our production editor has done thoroughly layout work for the reference. If any changes are necessary, please highlight them. Thanks for your cooperation. Please do not delete nonredundant ref. or add any new references. According to our requirements, no content changes are allowed after the article is accepted. Hope you can understand.    Author response: Thank you for doing this. As requested, we did not change the references format. As requested below, we removed one duplicated reference.
\newblock Deep learning in cancer pathology: A new generation of clinical
  biomarkers.
\newblock {\em Br. J. Cancer} {\bf 2021}, {\em 124},~686--696.

\bibitem[Zhu et~al.(2016)Zhu, Yao, and Huang]{zhu2016deep}
Zhu, X.; Yao, J.; Huang, J.
\newblock Deep convolutional neural network for survival analysis with
  pathological images.
\newblock In \emph{Proceedings of the 2016 IEEE International Conference on
  Bioinformatics and Biomedicine (BIBM)}; IEEE:  Piscataway, NJ, USA, %MDPI: We added the location of the publisher. Please confirm. the same below.   Author response: We confirm.
2016; pp.~544--547.

\bibitem[Orlov et~al.(2010)Orlov, Chen, Eckley, Macura, Shamir, Jaffe, and
  Goldberg]{orlov2010automatic}
Orlov, N.V.; Chen, W.W.; Eckley, D.M.; Macura, T.J.; Shamir, L.; Jaffe, E.S.;
  Goldberg, I.G.
\newblock Automatic classification of lymphoma images with transform-based
  global features.
\newblock {\em IEEE Trans. Inf. Technol. Biomed.} {\bf
  2010}, {\em 14},~1003--1013.

\bibitem[Xiao et~al.(2024)Xiao, Li, Yan, Gao, and Wang]{xiao2024convolutional}
Xiao, M.; Li, Y.; Yan, X.; Gao, M.; Wang, W.
\newblock Convolutional neural network classification of cancer cytopathology
  images: Taking breast cancer as an example.
\newblock In Proceedings of the 2024 7th International
  Conference on Machine Vision and Applications, Singapore , 12--14 March %MDPI: We added the location and date of the conference. Please confirm. the same below.    Author response: We confirm.
2024, pp. 145--149.

\bibitem[Xu et~al.(2025)Xu, Khan, Song, and Meijering]{xu2025edge}
Xu, Y.; Khan, T.M.; Song, Y.; Meijering, E.
\newblock Edge deep learning in computer vision and medical diagnostics: A
  comprehensive survey.
\newblock {\em Artif. Intell. Rev.} {\bf 2025}, {\em 58},~93.

\bibitem[Ciga et~al.(2022)Ciga, Xu, and Martel]{ciga2022self}
Ciga, O.; Xu, T.; Martel, A.L.
\newblock Self supervised contrastive learning for digital histopathology.
\newblock {\em Mach. Learn. Appl.} {\bf 2022}, {\em 7},~100198.

\bibitem[De~Matos et~al.(2021)De~Matos, Ataky, de~Souza Britto~Jr, Soares~de
  Oliveira, and Lameiras~Koerich]{de2021machine}
De~Matos, J.; Ataky, S.T.M.; de~Souza Britto~Jr, A.; Soares~de Oliveira, L.E.;
  Lameiras~Koerich, A.
\newblock Machine learning methods for histopathological image analysis: A
  review.
\newblock {\em Electronics} {\bf 2021}, {\em 10},~562.

\bibitem[Rajadurai et~al.(2024)Rajadurai, Perumal, Ijaz, and
  Chowdhary]{rajadurai2024precisionlymphonet}
Rajadurai, S.; Perumal, K.; Ijaz, M.F.; Chowdhary, C.L.
\newblock Precisionlymphonet: Advancing malignant lymphoma diagnosis via
  ensemble transfer learning with cnns.
\newblock {\em Diagnostics} {\bf 2024}, {\em 14},~469.

\bibitem[Chen et~al.(2024)Chen, Li, Li, Rahaman, Li, Wu, Sun, Grzegorzek, and
  Li]{chen2024can}
Chen, H.; Li, X.; Li, C.; Rahaman, M.M.; Li, X.; Wu, J.; Sun, H.; Grzegorzek,
  M.; Li, X.
\newblock What can machine vision do for lymphatic histopathology image
  analysis: A comprehensive review.
\newblock {\em Artif. Intell. Rev.} {\bf 2024}, {\em 57},~71.

\bibitem[Shen et~al.(2017)Shen, Wu, and Suk]{shen2017deep}
Shen, D.; Wu, G.; Suk, H.I.
\newblock Deep learning in medical image analysis.
\newblock {\em Annu. Rev. Biomed. Eng.} {\bf 2017}, {\em
  19},~221--248.

\bibitem[Irede et~al.(2024)Irede, Aworinde, Lekan, Amienghemhen, Okonkwo,
  Onivefu, and Ifijen]{irede2024medical}
Irede, E.L.; Aworinde, O.R.; Lekan, O.K.; Amienghemhen, O.D.; Okonkwo, T.P.;
  Onivefu, A.P.; Ifijen, I.H.
\newblock Medical imaging: A critical review on X-ray imaging for the detection
  of infection.
\newblock {\em Biomed. Mater. Devices} {\bf 2024}, \emph{4}%MDPI: We added the volume, please confirm.  Author response: We confirm.
, 1--45.

\bibitem[Weigel(2015)]{weigel2015extended}
Weigel, M.
\newblock Extended phase graphs: Dephasing, RF pulses, and echoes-pure and
  simple.
\newblock {\em J. Magn. Reson. Imaging} {\bf 2015}, {\em
  41},~266--295.

\bibitem[Haris et~al.(2015)Haris, Yadav, Rizwan, Singh, Wang, Hariharan, Reddy,
  and Marincola]{haris2015molecular}
Haris, M.; Yadav, S.K.; Rizwan, A.; Singh, A.; Wang, E.; Hariharan, H.; Reddy,
  R.; Marincola, F.M.
\newblock Molecular magnetic resonance imaging in cancer.
\newblock {\em J. Transl. Med.} {\bf 2015}, {\em 13},~313%MDPI: We revised the page, please confirm.  Author response: We confirm.
.

\bibitem[Schwenck et~al.(2023)Schwenck, Sonanini, Cotton, Rammensee,
  la~Foug{\`e}re, Zender, and Pichler]{schwenck2023advances}
Schwenck, J.; Sonanini, D.; Cotton, J.M.; Rammensee, H.G.; la~Foug{\`e}re, C.;
  Zender, L.; Pichler, B.J.
\newblock Advances in {PET} imaging of cancer.
\newblock {\em Nat. Rev. Cancer} {\bf 2023}, {\em 23},~474--490.

\bibitem[Reuveni et~al.(2011)Reuveni, Motiei, Romman, Popovtzer, and
  Popovtzer]{reuveni2011targeted}
Reuveni, T.; Motiei, M.; Romman, Z.; Popovtzer, A.; Popovtzer, R.
\newblock Targeted gold nanoparticles enable molecular {CT} imaging of cancer:
  an in vivo study.
\newblock {\em Int. J. Nanomed.} {\bf 2011}, \emph{6%MDPI: We added the volume, please confirm.  Author response: We confirm.
}, 2859--2864.

\bibitem[Ahmed et~al.(2020)Ahmed, Parvin, Haque, and Uddin]{ahmed2020lung}
Ahmed, T.; Parvin, M.S.; Haque, M.R.; Uddin, M.S.
\newblock Lung cancer detection using {CT} image based on {3D} convolutional
  neural network.
\newblock {\em J. Comput. Commun.} {\bf 2020}, {\em 8},~35.

\bibitem[Schwartz et~al.(2022)Schwartz, Sawyer, Thurston, Barton, and
  Ditzler]{schwartz2022ovarian}
Schwartz, D.; Sawyer, T.W.; Thurston, N.; Barton, J.; Ditzler, G.
\newblock Ovarian cancer detection using optical coherence tomography and
  convolutional neural networks.
\newblock {\em Neural Comput. Appl.} {\bf 2022}, {\em
  34},~8977--8987.

\bibitem[Ayana et~al.(2021)Ayana, Dese, and Choe]{ayana2021transfer}
Ayana, G.; Dese, K.; Choe, S.W.
\newblock Transfer learning in breast cancer diagnoses via ultrasound imaging.
\newblock {\em Cancers} {\bf 2021}, {\em 13},~738.

\bibitem[Islam et~al.(2022)Islam, Poly, Walther, Yeh, Seyed-Abdul, Li, and
  Lin]{islam2022deep}
Islam, M.M.; Poly, T.N.; Walther, B.A.; Yeh, C.Y.; Seyed-Abdul, S.; Li, Y.C.;
  Lin, M.C.
\newblock Deep learning for the diagnosis of esophageal cancer in endoscopic
  images: A systematic review and meta-analysis.
\newblock {\em Cancers} {\bf 2022}, {\em 14},~5996.

\bibitem[LeCun et~al.(2015)LeCun, Bengio, and Hinton]{lecun2015deep}
LeCun, Y.; Bengio, Y.; Hinton, G.
\newblock Deep learning.
\newblock {\em Nature} {\bf 2015}, {\em 521},~436--444.

\bibitem[Gu et~al.(2018)Gu, Wang, Kuen, Ma, Shahroudy, Shuai, Liu, Wang, Wang,
  and Cai]{gu2018recent}
Gu, J.; Wang, Z.; Kuen, J.; Ma, L.; Shahroudy, A.; Shuai, B.; Liu, T.; Wang,
  X.; Wang, G.; Cai, J.
\newblock Recent advances in convolutional neural networks.
\newblock {\em Pattern Recognit.} {\bf 2018}, {\em 77},~354--377.

\bibitem[Yamashita et~al.(2018)Yamashita, Nishio, Do, and
  Togashi]{yamashita2018convolutional}
Yamashita, R.; Nishio, M.; Do, R.K.G.; Togashi, K.
\newblock Convolutional neural networks: An overview and application in
  radiology.
\newblock {\em Insights Into Imaging} {\bf 2018}, {\em 9},~611--629.

\bibitem[Litjens et~al.(2017)Litjens, Kooi, Bejnordi, Setio, Ciompi,
  Ghafoorian, Van Der~Laak, Van~Ginneken, and S{\'a}nchez]{litjens2017survey}
Litjens, G.; Kooi, T.; Bejnordi, B.E.; Setio, A.A.A.; Ciompi, F.; Ghafoorian,
  M.; Van Der~Laak, J.A.; Van~Ginneken, B.; S{\'a}nchez, C.I.
\newblock A survey on deep learning in medical image analysis.
\newblock {\em Med. Image Anal.} {\bf 2017}, {\em 42},~60--88.

\bibitem[Anwar et~al.(2018)Anwar, Majid, Qayyum, Awais, Alnowami, and
  Khan]{anwar2018medical}
Anwar, S.M.; Majid, M.; Qayyum, A.; Awais, M.; Alnowami, M.; Khan, M.K.
\newblock Medical image analysis using convolutional neural networks: A review.
\newblock {\em J. Med. Syst.} {\bf 2018}, {\em 42},~1--13.

\bibitem[Chen et~al.(2017)Chen, Shi, Zhang, Wu, and Guizani]{chen2017deep}
Chen, M.; Shi, X.; Zhang, Y.; Wu, D.; Guizani, M.
\newblock Deep feature learning for medical image analysis with convolutional
  autoencoder neural network.
\newblock {\em IEEE Trans. Big Data} {\bf 2017}, {\em 7},~750--758.

\bibitem[Obermeyer et~al.(2019)Obermeyer, Powers, Vogeli, and
  Mullainathan]{obermeyer2019dissecting}
Obermeyer, Z.; Powers, B.; Vogeli, C.; Mullainathan, S.
\newblock Dissecting racial bias in an algorithm used to manage the health of
  populations.
\newblock {\em Science} {\bf 2019}, {\em 366},~447--453.

\bibitem[Dhar and Shamir(2021)]{dhar2021evaluation}
Dhar, S.; Shamir, L.
\newblock Evaluation of the benchmark datasets for testing the efficacy of deep
  convolutional neural networks.
\newblock {\em Vis. Inform.} {\bf 2021}, {\em 5},~92--101.

\bibitem[Ball(2023)]{ball2023ai}
Ball, P.
\newblock Is AI leading to a reproducibility crisis in science?
\newblock {\em Nature} {\bf 2023}, {\em 624},~22--25.

\bibitem[Torralba and Efros(2011)]{torralba2011unbiased}
Torralba, A.; Efros, A.A.
\newblock Unbiased look at dataset bias.
\newblock In \emph{Proceedings of the Computer Vision and Pattern Recognition 2011};
  IEEE: Piscataway, NJ, USA,  2011; pp. 1521--1528.

\bibitem[Ganganwar(2012)]{ganganwar2012overview}
Ganganwar, V.
\newblock An overview of classification algorithms for imbalanced datasets.
\newblock {\em Int. J. Emerg. Technol. Adv. Eng.} {\bf 2012}, {\em 2},~42--47.

\bibitem[Shamir(2008)]{shamir2008evaluation}
Shamir, L.
\newblock Evaluation of face datasets as tools for assessing the performance of
  face recognition methods.
\newblock {\em Int. J. Comput. Vis.} {\bf 2008}, {\em
  79},~225--230.

\bibitem[Erukude et~al.(2024)Erukude, Joshi, and
  Shamir]{erukude2024identifying}
Erukude, S.T.; Joshi, A.; Shamir, L.
\newblock Identifying Bias in Deep Neural Networks Using Image Transforms.
\newblock {\em Computers} {\bf 2024}, {\em 13},~341.

\bibitem[Dhar and Shamir(2022)]{dhar2022systematic}
Dhar, S.; Shamir, L.
\newblock Systematic biases when using deep neural networks for annotating
  large catalogs of astronomical images.
\newblock {\em Astron. Comput.} {\bf 2022}, {\em 38},~100545.

\bibitem[Shamir(2011)]{shamir2011assessing}
Shamir, L.
\newblock Assessing the efficacy of low-level image content descriptors for
  computer-based fluorescence microscopy image analysis.
\newblock {\em J. Microsc.} {\bf 2011}, {\em 243},~284--292.

\bibitem[DeGrave et~al.(2021)DeGrave, Janizek, and Lee]{degrave2021ai}
DeGrave, A.J.; Janizek, J.D.; Lee, S.I.
\newblock AI for radiographic COVID-19 detection selects shortcuts over signal.
\newblock {\em Nat. Mach. Intell.} {\bf 2021}, {\em 3},~610--619.

\bibitem[Zech et~al.(2018)Zech, Badgeley, Liu, Costa, Titano, and
  Oermann]{zech2018variable}
Zech, J.R.; Badgeley, M.A.; Liu, M.; Costa, A.B.; Titano, J.J.; Oermann, E.K.
\newblock Variable generalization performance of a deep learning model to
  detect pneumonia in chest radiographs: A cross-sectional study.
\newblock {\em PLoS Med.} {\bf 2018}, {\em 15},~e1002683.

\bibitem[Behar and Shrivastava(2022)]{behar2022resnet50}
Behar, N.; Shrivastava, M.
\newblock ResNet50-Based Effective Model for Breast Cancer Classification Using
  Histopathology Images.
\newblock {\em CMES---Comput. Model. Eng. Sci.} {\bf 2022},
  {\em 130}, 824--839. %MDPI: We adde the page, please confirm.

\bibitem[Pattanaik et~al.(2022)Pattanaik, Mishra, Siddique, Gopikrishna, and
  Satapathy]{pattanaik2022breast}
Pattanaik, R.K.; Mishra, S.; Siddique, M.; Gopikrishna, T.; Satapathy, S.
\newblock Breast cancer classification from mammogram images using extreme
  learning machine-based DenseNet121 model.
\newblock {\em J. Sens.} {\bf 2022}, {\em 2022},~2731364.

\bibitem[Al~Husaini et~al.(2021)Al~Husaini, Habaebi, Gunawan, Islam, and
  Hameed]{al2021automatic}
Al~Husaini, M.A.S.; Habaebi, M.H.; Gunawan, T.S.; Islam, M.R.; Hameed, S.A.
\newblock Automatic breast cancer detection using inception V3 in thermography.
\newblock In \emph{Proceedings of the 2021 8th International Conference on Computer
  and Communication Engineering (ICCCE)}; IEEE: Piscataway, NJ, USA,  2021; pp.~255--258.

\bibitem[Manasa and Murthy(2021)]{manasa2021skin}
Manasa, K.; Murthy, G.V.
\newblock Skin Cancer Detection Using VGG-16.
\newblock {\em Eur. J. Mol. Clin. Med.} {\bf 2021},
  {\em 8},~1419--1427.

\bibitem[{\c{C}}akmak and Pacal(2025)]{ccakmak2025deep}
{\c{C}}akmak, Y.; Pacal, N.
\newblock Deep learning for automated breast cancer detection in ultrasound: A
  comparative study of four cnn architectures.
\newblock {\em Artif. Intell. Appl. Sci.} {\bf 2025}, {\em
  1},~13--19.

\bibitem[Desai and Mahto(2025)]{desai2025multi}
Desai, A.; Mahto, R.
\newblock Multi-class classification of breast cancer subtypes using ResNet
  architectures on histopathological images.
\newblock {\em J. Imaging} {\bf 2025}, {\em 11},~284.

\bibitem[Emegano et~al.(2025)Emegano, Mustapha, Ozsahin, Ozsahin, and
  Uzun]{emegano2025histopathology}
Emegano, D.I.; Mustapha, M.T.; Ozsahin, D.U.; Ozsahin, I.; Uzun, B.
\newblock Histopathology-based prostate cancer classification using ResNet: A
  comprehensive deep learning analysis.
\newblock {\em J. Imaging Inform. Med.} {\bf 2025},  \emph{39}, 604--619 %MDPI: We revised the page and added volume, please confirm.   Author response: We confirm.
.

\bibitem[Jusman et~al.(2025)Jusman, Nurkholid, and
  Ramadani]{jusman2025comparison}
Jusman, Y.; Nurkholid, M.A.F.; Ramadani, D.A.
\newblock Comparison of Prostate Cell Image Classification Using CNN; Xception
  and DenseNet-201.
\newblock In \emph{Proceedings of the 2025 International Conference on Computer
  Sciences, Engineering, and Technology Innovation (ICoCSETI)}; IEEE: Piscataway, NJ, USA,  2025; pp.~25--29.

\bibitem[Siahaan and Sianipar(2021)]{siahaan2021implementasi}
Siahaan, V.; Sianipar, R.H.
\newblock {\em Implementasi OpenCV dan PIL (Python Imaging Library) dengan
  Python/MySQL}; Balige Publishing:  Balige, Indonesia, 2021.

\bibitem[Kingma and Ba(2014)]{kingma2014adam}
Kingma, D.P.; Ba, J.
\newblock Adam: A method for stochastic optimization.
\newblock {\em arXiv} {\bf 2014}, arXiv:1412.6980.

\bibitem[Wilson and Martinez(2001)]{wilson2001need}
Wilson, D.R.; Martinez, T.R.
\newblock The need for small learning rates on large problems.
\newblock In \emph{Proceedings of the IJCNN'01. International Joint Conference on
  Neural Networks. Proceedings (Cat. No. 01CH37222)}; IEEE: Piscataway, NJ, USA,  2001; Volume~1, pp.~115--119.

\bibitem[Hutson et~al.(2019)Hutson, Andresen, Tygart, and
  Turner]{hutson2019managing}
Hutson, K.; Andresen, D.; Tygart, A.; Turner, D.
\newblock Managing a heterogeneous cluster. In {\em Practice and Experience in
  Advanced Research Computing 2019: Rise of the Machines (Learning)};  Association for Computing Machinery: Chicago, IL, USA, 2019; %MDPI: We added the publisher name and location, please confirm.   Author response: We confirm.
  pp. 1--6.

\bibitem[Yang et~al.(2023{\natexlab{a}})Yang, Shi, Wei, Liu, Zhao, Ke, Pfister,
  and Ni]{medmnistv2}
Yang, J.; %MDPI: Refs. 49 and 50 are duplicated. Please remove duplicated references and rearrange all the references to appear in numerical order. Please ensure that there are no duplicated references.   Author response: The duplicated reference was removed. We ensure that the correct reference is cited to avoid "?" instead of the citations.
 Shi, R.; Wei, D.; Liu, Z.; Zhao, L.; Ke, B.; Pfister, H.; Ni, B.
\newblock MedMNIST v2-A large-scale lightweight benchmark for 2D and 3D
  biomedical image classification.
\newblock {\em Sci. Data} {\bf 2023}, {\em 10},~41.

%\bibitem[Yang et~al.(2023{\natexlab{b}})Yang, Shi, Wei, Liu, Zhao, Ke, Pfister,
%  and Ni]{yang2023medmnist}
%Yang, J.; Shi, R.; Wei, D.; Liu, Z.; Zhao, L.; Ke, B.; Pfister, H.; Ni, B.
%\newblock Medmnist v2-a large-scale lightweight benchmark for {2D} and {3D}
 % biomedical image classification.
%\newblock {\em Sci. Data} {\bf 2023}, {\em 10},~41.

\bibitem[Doerrich et~al.(2025)Doerrich, Di~Salvo, Brockmann, and
  Ledig]{doerrich2025rethinking}
Doerrich, S.; Di~Salvo, F.; Brockmann, J.; Ledig, C.
\newblock Rethinking model prototyping through the MedMNIST+ dataset
  collection.
\newblock {\em Sci. Rep.} {\bf 2025}, {\em 15},~7669.

\bibitem[Kather et~al.(2019)Kather, Krisam, Charoentong, Luedde, Herpel, Weis,
  Gaiser, Marx, Valous, and Ferber]{kather2019predicting}
Kather, J.N.; Krisam, J.; Charoentong, P.; Luedde, T.; Herpel, E.; Weis, C.A.;
  Gaiser, T.; Marx, A.; Valous, N.A.; Ferber, D.
\newblock Predicting survival from colorectal cancer histology slides using
  deep learning: A retrospective multicenter study.
\newblock {\em PLoS Med.} {\bf 2019}, {\em 16},~e1002730.

\bibitem[Tschandl et~al.(2018)Tschandl, Rosendahl, and
  Kittler]{tschandl2018ham10000}
Tschandl, P.; Rosendahl, C.; Kittler, H.
\newblock The HAM10000 dataset, a large collection of multi-source
  dermatoscopic images of common pigmented skin lesions.
\newblock {\em Sci. Data} {\bf 2018}, {\em 5},~180161 %MDPI: We revised the page number,please confirm.  Author response: We conform.
.

\bibitem[Al-Dhabyani et~al.(2020)Al-Dhabyani, Gomaa, Khaled, and
  Fahmy]{al2020dataset}
Al-Dhabyani, W.; Gomaa, M.; Khaled, H.; Fahmy, A.
\newblock Dataset of breast ultrasound images.
\newblock {\em Data Brief} {\bf 2020}, {\em 28},~104863.

\bibitem[Samuel(2011)]{samuel2011lung}
Samuel, G.
\newblock The Lung Image Database Consortium (LIDC) and Image Database resource
  initiative (IDRI): A completed reference database of lung nodules on CT
  scans.
\newblock {\em Med. Phys.} {\bf 2011}, {\em 38},~2.

\bibitem[Spanhol et~al.(2015)Spanhol, Oliveira, Petitjean, and
  Heutte]{spanhol2015dataset}
Spanhol, F.A.; Oliveira, L.S.; Petitjean, C.; Heutte, L.
\newblock A dataset for breast cancer histopathological image classification.
\newblock {\em {IEEE} Trans. Biomed. Eng.} {\bf 2015}, {\em
  63},~1455--1462.

\bibitem[Gutman et~al.(2016)Gutman, Codella, Celebi, Helba, Marchetti, Mishra,
  and Halpern]{gutman2016skin}
Gutman, D.; Codella, N.C.; Celebi, E.; Helba, B.; Marchetti, M.; Mishra, N.;
  Halpern, A.
\newblock Skin lesion analysis toward melanoma detection: A challenge at the
  international symposium on biomedical imaging (ISBI) 2016, hosted by the
  international skin imaging collaboration (ISIC).
\newblock {\em arXiv} {\bf 2016}, arXiv:1605.01397.

\bibitem[Codella et~al.(2018)Codella, Gutman, Celebi, Helba, Marchetti, Dusza,
  Kalloo, Liopyris, Mishra, and Kittler]{codella2018skin}
Codella, N.C.; Gutman, D.; Celebi, M.E.; Helba, B.; Marchetti, M.A.; Dusza,
  S.W.; Kalloo, A.; Liopyris, K.; Mishra, N.; Kittler, H.
\newblock Skin lesion analysis toward melanoma detection: A challenge at the
  2017 international symposium on biomedical imaging (isbi), hosted by the
  international skin imaging collaboration (isic).
\newblock In \emph{Proceedings of the 2018 IEEE 15th International Symposium on
  Biomedical Imaging (ISBI 2018)}; IEEE: Piscataway, NJ, USA,  2018, pp. 168--172.

\bibitem[Combalia et~al.(2019)Combalia, Codella, Rotemberg, Helba, Vilaplana,
  Reiter, Carrera, Barreiro, Halpern, and Puig]{combalia2019bcn20000}
Combalia, M.; Codella, N.C.; Rotemberg, V.; Helba, B.; Vilaplana, V.; Reiter,
  O.; Carrera, C.; Barreiro, A.; Halpern, A.C.; Puig, S.
\newblock Bcn20000: Dermoscopic lesions in the wild.
\newblock {\em arXiv} {\bf 2019}, arXiv:1908.02288.

\bibitem[Abdallah et~al.(2023)Abdallah, Marion, Tauber, Carlier, Hatt, and
  Chauvet]{abdallah2023enhancing}
Abdallah, N.; Marion, J.M.; Tauber, C.; Carlier, T.; Hatt, M.; Chauvet, P.
\newblock Enhancing histopathological image classification of invasive ductal
  carcinoma using hybrid harmonization techniques.
\newblock {\em Sci. Rep.} {\bf 2023}, {\em 13},~20014.

\bibitem[Janowczyk and Madabhushi(2016)]{Janowczyk2016DeepLearningDP}
Janowczyk, A.; Madabhushi, A.
\newblock Deep learning for digital pathology image analysis: A comprehensive
  tutorial with selected use cases.
\newblock {\em J. Pathol. Inform.} {\bf 2016}, {\em 7},~29.
\newblock {\url{https://doi.org/10.4103/2153-3539.186902}}.

\end{thebibliography}
\end{document}